\documentclass[acmsmall,screen]{acmart}

\AtBeginDocument{%
  }

\acmISBN{978-1-4503-XXXX-X/18/06}
\usepackage{natbib}
\usepackage[ruled,linesnumbered]{algorithm2e}
\usepackage[normalem]{ulem}
\usepackage{pifont} 
\usepackage{multirow}
\usepackage{subfigure}
\usepackage{makecell}
\usepackage{tcolorbox}
\usepackage{hyperref}
\makeatletter
\def\UrlAlphabet{%
	\do\a\do\b\do\c\do\d\do\e\do\f\do\g\do\h\do\i\do\j%
	\do\k\do\l\do\m\do\n\do\o\do\p\do\q\do\r\do\s\do\t%
	\do\u\do\v\do\w\do\x\do\y\do\z\do\A\do\B\do\C\do\D%
	\do\E\do\F\do\G\do\H\do\I\do\J\do\K\do\L\do\M\do\N%
	\do\O\do\P\do\Q\do\R\do\S\do\T\do\U\do\V\do\W\do\X%
	\do\Y\do\Z}
\def\UrlDigits{\do\1\do\2\do\3\do\4\do\5\do\6\do\7\do\8\do\9\do\0}
\g@addto@macro{\UrlBreaks}{\UrlOrds}
\g@addto@macro{\UrlBreaks}{\UrlAlphabet}
\g@addto@macro{\UrlBreaks}{\UrlDigits}

\begin{document}

\title{An Effective Docker Image Slimming Approach Based on Source Code Data Dependency Analysis}


\author{Jiaxuan Han}
\email{zhanSxDrive30i@gmail.com}
\orcid{0009-0008-9963-3997}
\author{Cheng Huang}
\email{opcodesec@gmail.com}
\orcid{0000-0002-5871-946X}
\authornote{Corresponding Author.}
\author{Jiayong Liu}
\email{ljy@scu.edu.cn}
\affiliation{%
	\department{School of Cyber Science and Engineering}
	\institution{Sichuan University}
	\city{Chengdu}
	\state{Sichuan}
	\country{China}
}

\author{Tianwei Zhang}
\email{tianwei.zhang@ntu.edu.sg}
\orcid{0000-0001-6595-6650}
\affiliation{%
	\department{College of Computing and Data Science}
	\institution{Nanyang Technological University}
	\city{Singapore}
	\country{Singapore}
}

\renewcommand{\shortauthors}{Han et al.}

\begin{abstract}
Containerization is the mainstream of current software development, which enables software to be used across platforms without additional configuration of running environment. However, many images created by developers are redundant and contain unnecessary code, packages, and components. This excess not only leads to bloated images that are cumbersome to transmit and store but also increases the attack surface, making them more vulnerable to security threats. Therefore, image slimming has emerged as a significant area of interest. Nevertheless, existing image slimming technologies face challenges, particularly regarding the incomplete extraction of environment dependencies required by project code. In this paper, we present a novel image slimming model named $\delta$-SCALPEL. This model employs static data dependency analysis to extract the environment dependencies of the project code and utilizes a data structure called the command linked list for modeling the image's file system. We select 20 NPM projects and two official Docker Hub images to construct a dataset for evaluating $\delta$-SCALPEL. The evaluation results show that $\delta$-SCALPEL can reduce image sizes by up to 61.4\% while ensuring the normal operation of these projects.
\end{abstract}

\ccsdesc[500]{Security and privacy~File system security}
\ccsdesc[500]{Software and its engineering~File systems management}
\ccsdesc[500]{Software and its engineering~Automated static analysis}

\keywords{Docker image, Image slimming, Static code analysis, Data dependency analysis, Command linked list}


\maketitle

\section{Introduction}

As a lightweight virtualization technology designed to create isolated environments, containers differ from virtual machines (VMs) by relying on process-level isolation rather than operating system-level resource isolation \cite{soltesz2007container, xavier2013performance, morabito2015hypervisors}. Docker \cite{merkel2014docker}, as a mainstream tool for creating containers, has developed rapidly in recent years. Its advantage lies in allowing developers to package various applications and their dependencies into Docker images, which can then be installed and run on any physical device, such as Linux or Windows devices, to achieve virtualization. This allows applications to be completely decoupled from the underlying hardware, enabling flexible migration and deployment between physical machines \cite{zou2019docker,combe2016docker}. As a result, engineers are freed from complex environment configurations, significantly improving the efficiency and reducing potential risks during deployment. \cite{muhtaroglu2017testing}.

Docker Hub\footnote{\url{https://hub.docker.com/}} is one of the most popular Docker image registries. Similar to open-source package repositories like NPM\footnote{\url{https://www.npmjs.com/}}, Maven\footnote{\url{https://mvnrepository.com/}}, and PyPI\footnote{\url{https://pypi.org/}}, it provides a centralized platform where users can access, share, and distribute Docker images published by developers \cite{liu2020understanding, zhao2019large, lin2020large}. In the software development process, developers can define the image-building process using a Dockerfile and specify the base image with the \texttt{FROM} instruction. During the image building, Docker downloads the specified base image from Docker Hub and then builds the image according to the user's requirements based on that base image \cite{henkel2020dataset,eng2021revisiting}. 

\begin{figure}[!h]
	\centering
	\includegraphics[width=0.75\linewidth]{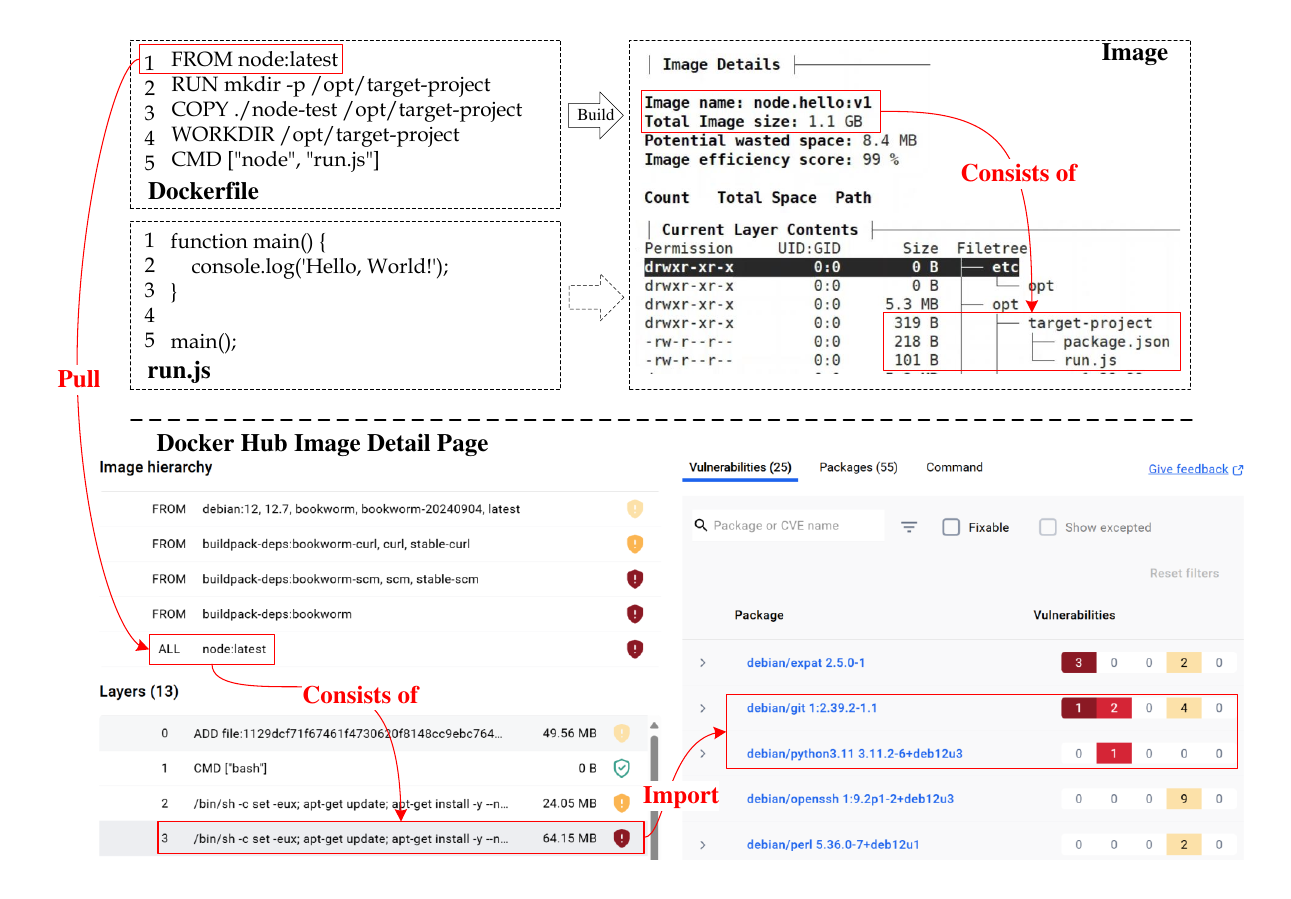}
	\caption{An example of security risk caused by image resource redundancy.}\label{example-image-build}
\end{figure}

Although users can extend a base image to build one that meets the project’s operational requirements, the resulting image may include redundant resources, wasting server storage space and potentially introducing security risks \cite{chainguard2024}. For example, in Fig. \ref{example-image-build}, a Dockerfile is used to extend the base image \texttt{node:latest} as the environment for the \texttt{run.js} file. We use the Dive\footnote{\url{https://github.com/wagoodman/dive}} tool to inspect the built image and find that it mainly contains two parts: the base environment part and the project code part. The base environment part takes up 1.1GB, while the project code part only occupies 319 B. The inspection result shows that there is a great waste of resources in the image extended from the base image. At the same time, we analyze the detail page\footnote{\url{https://hub.docker.com/layers/library/node/latest/images/sha256-eb5bb667442cadcd1bb8e6b3d44b2d11bbc5beb280db7f022872d33177b61ca1?context=explore}} of the \texttt{node:latest} image on Docker Hub and find that it introduces vulnerable \texttt{Git} and \texttt{Python} environments in the third layer, even though these environments are not required for the project code. Therefore, it is necessary to slim the image to reduce storage waste and minimize the attack surface.

\begin{figure}[!b]
	\centering
	\includegraphics[width=0.6\linewidth]{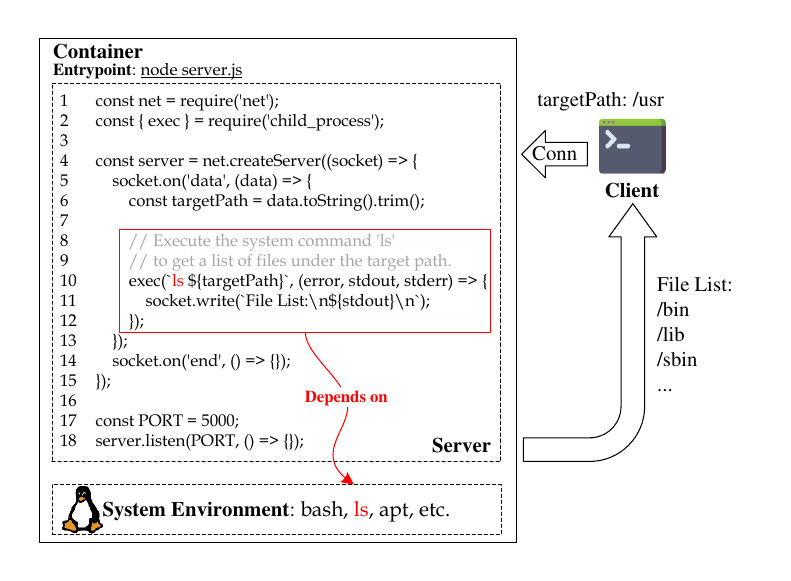}
	\caption{An example that the code needs to depend on the system environment.}\label{shortcoming-1}
\end{figure}

The key of image slimming is how to determine the necessary environment for the project code. Slim is a popular tool for reducing the Docker image size \cite{slimtoolkit2024}. It creates a temporary container for the target image and hooks key files in the container's file system to identify the necessary environment for running the container. This allows for the exclusion of content irrelevant to the project code, enabling efficient image slimming. The advantage of Slim is that it can reduce the image size by up to 30 times without altering anything in the image. However, Slim also exhibits two limitations: 

\noindent
\textbf{\uline{Limitation 1: The extraction of the necessary environment for the project code operation is incomplete.}} Since Slim determines the required system environment based on the runtime behavior of the project code, like many dynamic program analysis techniques, it still faces the issue of incomplete code coverage \cite{Jiaxuan2022Vuld,dawoud2021bringing,ahmad2020stadart,ghavamnia2020confine}. Fig. \ref{shortcoming-1} is an example that the code needs to depend on the system environment. When the server is running, it first starts listening. Upon receiving a client connection, it calls the \texttt{exec} API (Application Programming Interface) to execute the system command \texttt{ls} that retrieves the file list from the folder specified by the client and returns the results. When using the Slim tool to slim this container, since there is no client connection, the code from lines 5 to 14 will not be executed. As a result, the project's dependency on the system command \texttt{ls} cannot be detected, leading to the project code not running properly in the slimmed container.

\begin{figure}[!ht]
	\centering
	\includegraphics[width=0.6\linewidth]{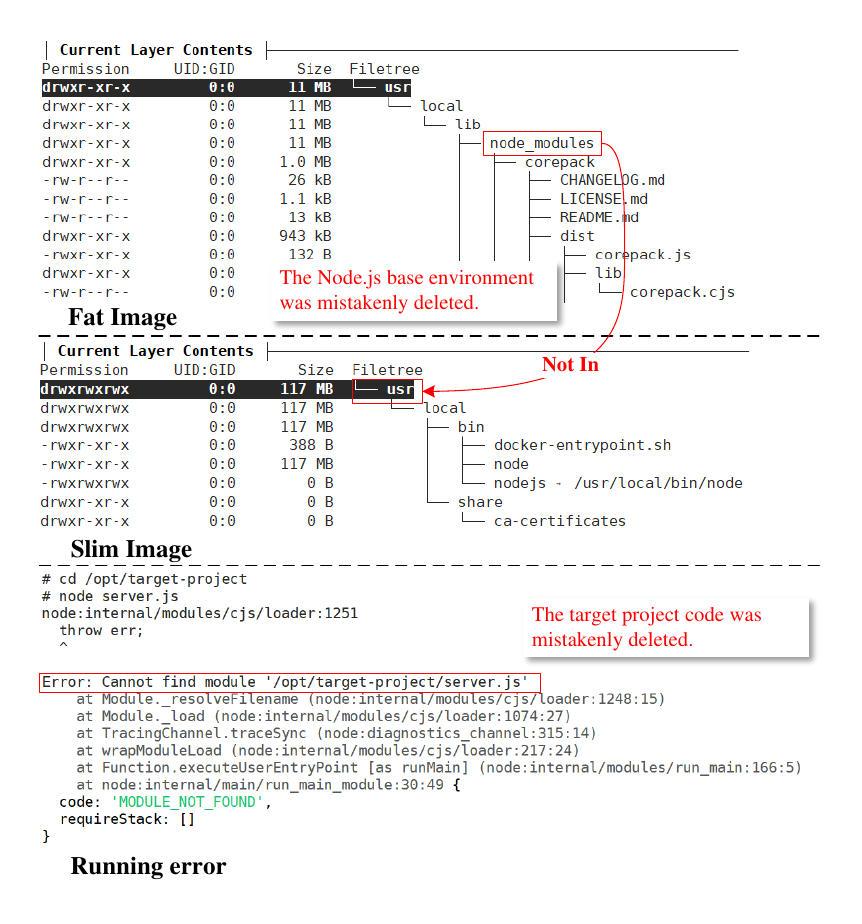}
	\vspace{-10pt}
	\caption{An example of project code operation failure caused by image slimming using Slim.}\label{shortcoming-2}
\end{figure}

\noindent
\textbf{\uline{Limitation 2: It is necessary to explicitly give the entry point of the image.}} This limitation is also due to Slim's reliance on the project's runtime behavior. As shown in Fig. \ref{shortcoming-2}, when the image is built, the commands to be executed at runtime (i.e., the image entry point) are not explicitly specified. As a result, Slim cannot identify the system environment required by the project code, leading to the erroneous removal of the Node.js base environment and the project code, which affects the normal operation of the container.

In order to solve these two limitations, we adopt a completely different strategy from Slim, using static code analysis technology to extract the environment dependencies of the project code. Specifically, we propose a novel Docker image slimming model, $\delta$-SCALPEL. $\delta$-SCALPEL extracts environment dependencies through static data dependency analysis of the project source code and its dependent software packages, determining the scope for image slimming. Simultaneously, it constructs a data structure called the command linked list to model the image's file system, which enhances the accuracy of extracting environment dependencies. Our model can slim the image whether the entry point is explicitly specified or not, while ensuring the normal operation of the project code. 

The major contributions of this paper are summarized as follows:

\begin{itemize}
	\item We propose $\delta$-SCALPEL, a novel Docker image slimming model. By utilizing static code analysis, this model overcomes the limitations of the Slim tool, which relies on runtime behavior and often results in unusable slimmed images. To the best of our knowledge, $\delta$-SCALPEL is the first Docker image slimming model that leverages static code analysis.
	
	\item We introduce a novel data structure called the command linked list. By scanning the image's file system, it models the links and dependencies between system commands and binary executable files, thereby ensuring the accuracy of image slimming.
	
	\item We conduct comprehensive evaluations of $\delta$-SCALPEL to assess its effectiveness and practical significance. The evaluation results show that $\delta$-SCALPEL can reduce the image size by 61.4\% at most, while ensuring that the slimmed image remains functional.
\end{itemize}

\section{Related Work}
\subsection{Static Code Analysis}

Static source code analysis is widely used in software engineering and security fields. Currently, it can be categorized into learning-based static analysis and traditional static analysis \cite{Jiaxuan2022Vuld}:

\noindent\textbf{\uline{Learning-based static analysis.}} In recent years, deep learning models for source code analysis have shown great promise. Han et al. \cite{han2024bjcnet,han2024bjenet,han2023bjxnet} proposed a series of methods for analyzing defects in Java source code. By converting the source code into specific intermediate representations, such as code property graph and code summary, deep learning models were applied to analyze these representations. Chen et al. \cite{chen2024egfe} designed an end-to-end grouping model for UI (User interface) design fragment elements based on multimodal learning, utilizing a Transformer encoder to model the relationships between UI elements. Ding et al. \cite{ding2023concord} proposed a new scheme for enhancing clone-aware data, designing multiple code transformation heuristics to mimic the cloning behavior of human developers. The original program and its clone variants are encoded with similar embeddings, while defective variants are distinguished with significantly different embeddings.

\noindent\textbf{\uline{Traditional static analysis.}} Traditional static analysis primarily relies on control flow analysis, data flow analysis, etc., to model source code, analyzing its behavior through a series of expert-defined rules. Huang et al. \cite{huang2024donapi} proposed a detection method for malicious NPM packages based on behavior sequence knowledge mapping. This method involves parsing the source code into the AST (Abstract Syntax Tree) and extracting the API call sequence through control flow and data dependency analysis. Currently, the most advanced learning-based static analysis methods predominantly utilize token-based Transformer models; however, these are not the most effective for capturing the code semantics required for vulnerability detection. Traditional static analysis methods, such as data flow analysis, can detect various types of bugs based on their root causes. Therefore, Steenhoek et al. \cite{steenhoek2024dataflow} combined data flow analysis with graph learning techniques to encode variable definitions and usages using effective bit-vector representations of data flow facts. Graph learning was then employed to propagate and aggregate data flow information, thereby simulating data flow calculations in data flow analysis. Wang et al. \cite{wang2023taintmini} expanded the JavaScript object dependence graph to account for the interactions between JavaScript and the webview layer, establishing a general data flow graph to trace data flow across four domains: the flow between the webview layer and the logic layer (i.e., JavaScript), the flow between asynchronous event handlers, the flow between different pages in the same mini-application, and the flow between different mini-applications.

\subsection{Docker Security}

Docker is a widely used technology for containerizing applications along with their dependencies, creating reproducible environments \cite{opdebeeck2023docker,martin2018docker}. Research on Docker security can be categorized into image \& container security and ecosystem security:

\noindent\textbf{\uline{Image \& container security.}} Image and container security primarily refers to the security of the system and software bundled within the container generated by an image, i.e., application security \cite{sultan2019container}. The container is deployed under the guidance of an automated deployment chain \cite{dockerHubBuilds}, which usually contains third-party programs, software packages, and components. Introducing these third-party elements may bring security risks to the container \cite{martin2018docker}. Tak et al. \cite{tak2018security} pointed out that packages included in images, such as perl, curl, and wget, may contain vulnerabilities. Therefore, software vulnerabilities, configuration defects, and malware are potential threat scenarios within this category \cite{souppaya2017application,lin2018measurement,javed2021understanding}.

\noindent\textbf{\uline{Ecosystem security.}} Docker Hub is a central registry where developers can obtain and push images. In a detailed study of Docker Hub images, Tarun Desikan et al. \cite{banyanSecurity} found that over 30\% of images in the official repository were highly susceptible to various security attacks. For unofficial images, i.e., those pushed by users without any formal verification from an authoritative entity, this figure rises to approximately 40\%. They highlighted that many official Docker Hub images contain packages with CVE (Common Vulnerabilities and Exposures) vulnerabilities, which are often unnecessary in certain cases. If not explicitly removed from the container, these packages may leave the container vulnerable to malicious attacks. Malicious images are also a key concern in the security of the Docker ecosystem. Spring et al. \cite{threatpostDocker} revealed that 17 malicious images hosted on Docker Hub allowed hackers to earn \$90,000 through cryptojacking, with these images being downloaded over 5 million times in a single year. The inheritance of Docker images can propagate security vulnerabilities from parent images to child images, thereby impacting the entire Docker ecosystem \cite{opdebeeck2023docker}.

\section{Methodology}

\begin{figure*}[!h]
	\centering
	\includegraphics[width=\linewidth]{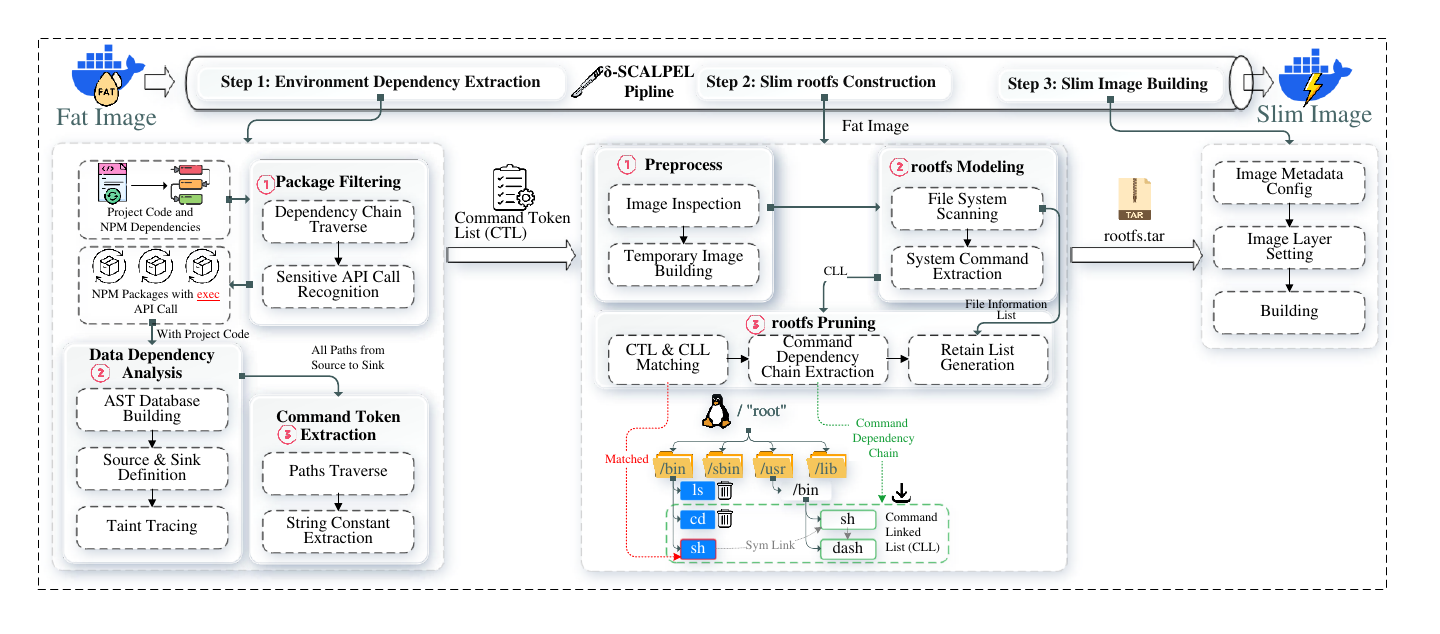}
	\caption{The framework of $\delta$-SCALPEL.}\label{framework}
\end{figure*}

\subsection{Overview}

The framework of $\delta$-SCALPEL is shown in Fig. \ref{framework}, which is divided into three steps:

\begin{itemize}
	\item \textbf{Step 1: Environment Dependency Extraction.} Create an image for data dependency analysis of the project code and its NPM dependencies, extracting environment dependencies (i.e., system commands executed during \texttt{exec} API calls) from the source code.
	
	\item \textbf{Step 2: Slim rootfs Construction.} Perform a static scan of the fat image to extract metadata and build a temporary image. Next, scan its file system, identify system commands, and determine which content should be removed based on the command token list extracted in step 1. Finally, construct a slim root file system (rootfs).
	
	\item \textbf{Step 3: Slim Image Building.} Create a slim image based on the \texttt{rootfs.tar} file constructed in step 2.
\end{itemize}

\subsection{Environment Dependency Extraction}

This step is divided into three modules: package filtering, data dependency analysis, and command token extraction. Before performing these modules, environment preparation is required. We choose Node.js as the default language and perform static code analysis using CodeQL\footnote{\url{https://github.com/github/codeql}}. In order to extract environment dependencies, we configure an image through a Dockerfile. Specifically, we set the base image to \texttt{node:current-slim}, and create the installation directory for NPM packages. Then, we configure the environment for CodeQL. Afterward, we initialize the target project with \texttt{npm install} command to install all its dependencies. Finally, an image for environment dependency extraction is generated.

\subsubsection{Package Filtering} 

With the expansion of the software ecosystem, developers increasingly import open source packages to enhance development efficiency. Consequently, during code execution, both control flow and data flow extend into these imported packages. So environment dependency extraction is required for these packages. However, analyzing the entire dependency chain of the project code is challenging due to the large number of packages involved. For example, in the nodejs-websocket\footnote{\url{https://www.npmjs.com/package/nodejs-websocket}} project, the dependency chain includes 30 packages. Analyzing data dependencies of these packages individually would be highly time-consuming. To optimize this process, we filter the packages in the project's dependency chain. In particular, we traverse the project's dependency chain, analyze the JavaScript code of each NPM package, and use regular expressions to identify packages with \texttt{exec} API calls. In this way, we can filter out most packages unrelated to environment dependency extraction, thereby speeding up the execution of $\delta$-SCALPEL.

\subsubsection{Data Dependency Analysis}

This is the core module of step 1, focusing on extracting string constants from \texttt{exec} API calls. Since developers do not always use string constants (i.e., the commands need to be executed) as parameters when calling the \texttt{exec} API, it is necessary to analyze data dependencies of the \texttt{exec} API calls. In this paper, we utilize CodeQL to achieve this. First, we use CodeQL to parse the target project and the NPM packages containing \texttt{exec} API calls, and build the AST database. Then, we define the source and sink nodes in the source code: sources include \texttt{VariableDeclarator}, \texttt{Assignment}, and \texttt{CallExpr} nodes, while sinks are the arguments of the \texttt{CallExpr} nodes named \texttt{exec}. Finally, we use the \texttt{TaintTracking} package of CodeQL to identify all paths from sources to sinks.

\subsubsection{Command Token Extraction}\label{cte} 

After obtaining all the paths from sources to sinks, we traverse the nodes along these paths to identify string constants and split them by spaces to extract command tokens. Additionally, the shell script files also have dependencies on the system environment. Therefore, we analyze all shell scripts in the project code folder and the \texttt{docker-entrypoint.sh} file, extracting tokens from them.

It should be noted that these tokens are not necessarily commands; they may also represent strings without command-related meaning, or command parameters. They are merely considered potential commands, and will be addressed further in Sect. \ref{rp}.

\subsection{Slim rootfs Construction}

After completing the extraction of environment dependencies, the command token list (CTL) is generated. The next step is to construct a slim rootfs for the fat image. In this step, we design three modules: preprocess, rootfs modeling, and rootfs pruning, which are used for analyzing the fat image's file system, identifying the content that needs to be removed, and constructing a slim rootfs.

\subsubsection{Preprocess}

First, we perform a static inspection of the fat image to obtain metadata, including its configuration and architecture. Then, we construct a temporary image based on the metadata, which contains a component called image analyzer. The goal of the image analyzer is to model the fat image's rootfs (rootfs modeling, Sect. \ref{rm}) and prune it (rootfs pruning, Sect. \ref{rp}) to create a slim rootfs.

\subsubsection{rootfs Modeling}\label{rm}

In this module, we first collect the file information list by scanning the entire rootfs to retrieve all files, folders, and their permissions. At the same time, we extract the symbolic soft link relationships between these files and folders. Then we extract all the commands supported by the system to build the command linked list (CLL).

The usage of commands can be categorized into two types: directly using the command name and using the absolute path of the command's binary file. So, we need to get the absolute path of a command. However, due to the symbolic soft link mechanism, one path can be linked to another, meaning the absolute path of a binary file obtained by the \texttt{which} command may not reflect its actual storage location. For example, running the \texttt{which sh} command returns the absolute path \texttt{/usr/bin/sh}. However, this path is a symbolic soft link to the \texttt{dash} command, whose absolute path is \texttt{/usr/bin/dash}. Additionally, \texttt{/bin} is a symbolic soft link to \texttt{/usr/bin}, meaning \texttt{/bin/sh} and \texttt{/usr/bin/sh} point to the same file. These indicate that for the two different ways of using commands, the paths that need to be included when generating the retain list are different. When the project code directly uses the \texttt{sh} command, the paths included in the retain list must be \texttt{[/usr/bin/sh, /usr/bin/dash]}, just as when \texttt{/usr/bin/sh} is used. However, when \texttt{/bin/sh} is used, the paths included in the retain list must be \texttt{[/bin/sh, /usr/bin/sh, /usr/bin/dash]}.

\begin{figure}[!htpb]
	\centering
	\includegraphics[width=0.7\linewidth]{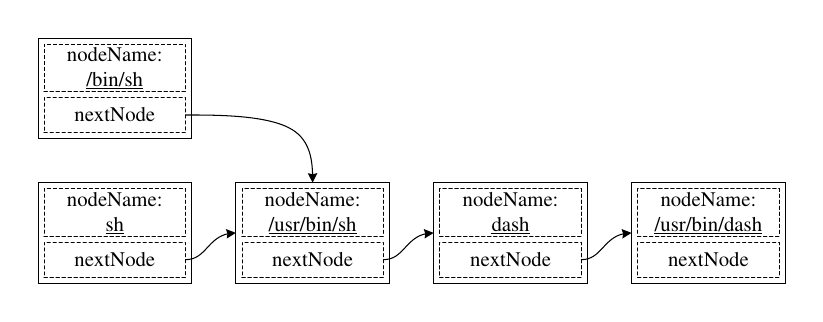}
	\caption{A command linked list example for the \texttt{sh} command.}\label{cmdlinklist}
\end{figure}

To this end, we design a novel data structure called command linked list to handle the above situation. First, we construct the command linked list using Algo. \ref{cllc}. Then, we extend it with Algo. \ref{clle} to handle the case of folder symbolic soft links. For the \texttt{sh} command, we can construct a command linked list as illustrated in Fig. \ref{cmdlinklist}.

\begin{minipage}[t]{0.48\textwidth}
	\begin{algorithm}[H]\tiny
		\SetAlgoLined
		\caption{CLL Construction}\label{cllc}
		\KwIn{Commands supported by the system $sysCmdList$}
		\KwResult{Command linked list $cmdLinkedList$}
		
		\SetKwFunction{MyFunction}{buildCmdLinkedList} 
		\SetKwProg{Fn}{func}{:}{\KwRet} 
		
		\Fn{\MyFunction{linkNode}}{
			
			\If{isCmdName(linkNode.nodeName)}{
				
				cmdAbsPath $\leftarrow$ getAbsPath(linkNode.nodeName);
				
			}
			\ElseIf{isSymLink(linkNode.nodeName)}{
				
				cmdAbsPath $\leftarrow$ getRefPath(linkNode.nodeName);
				
			}
			
			\If{cmdLinkedList.hasKey(cmdAbsPath)}{
				
				currentNode $\leftarrow$ cmdLinkedList.get(cmdAbsPath);
				
			} 
			\Else{
				
				currentNode $\leftarrow$ new(
				
				\Indp	nodeName=cmdAbsPath,
				
				nextNode=null
				
				\Indm	);
				
				cmdLinkedList.append(currentNode);
				
			}
			
			\If{linkNode.nextNode == null}{
				
				linkNode.nextNode $\leftarrow$ currentNode;
				
			}
			
			\MyFunction(linkNode.nextNode);
			
		}
		
		init(cmdLinkedList); \tcp{global variable}
		
		\For{sysCmd in sysCmdList}{
			
			\If{isBuiltInCmd(sysCmd)}{
				
				continue;
				
			}

			linkNode $\leftarrow$ new(
			
			\Indp nodeName=sysCmd, 
			
			nextNode=null
			
			\Indm );
			
			\MyFunction(linkNode);

		}
		
	\end{algorithm}
\end{minipage}
\hfill
\begin{minipage}[t]{0.48\textwidth}
	\begin{algorithm}[H]\tiny
		\SetAlgoLined
		\caption{CLL Expanding}\label{clle}
		\KwIn{Command Linked List $cmdLinkedList$}
		\KwResult{Command Linked List $cmdLinkedList$}
		
		\SetKwFunction{MyFunction}{buildCmdLinkedList}
		
		\For{linkNode in cmdLinkedList}{
			
			\If{!isPath(linkNode.name)}{
				continue;
			}
			
			allParentDirs $\leftarrow$ getAllParentDirs(linkNode.nodeName);
			
			\For{dir in allParentDirs}{
				
				isSymLink $\leftarrow$ false;
				
				hasSymLink $\leftarrow$ false;
				
				\If{isSymPath(dir)}{
					
					isSymLink $\leftarrow$ true;
					
					\tcc{dir is a symbolic soft link}
					
					refDir $\leftarrow$ getRefPath(dir);
					
					newCmdPath $\leftarrow$ linkNode.nodeName.replace(dir, refDir);
					
				}
				\ElseIf{hasSymPath(dir)}{
					
					\tcc{dir is not a symbolic soft link, but it has related symbolic soft link}
					
					hasSymLink $\leftarrow$ true;
					
					linkDir $\leftarrow$ getLinkPath(dir);
					
					newCmdPath $\leftarrow$ linkNode.nodeName.replace(dir, linkDir);		
				}
				
				\If{hasFile(newCmdPath)}{
					
					\If{!cmdLinkedList.hasKey(newCmdPath)}{
						
						newNode $\leftarrow$ new(
						
						\Indp nodeName=newCmdPath, 
						
						nextNode=null
						
						\Indm );
						
						\MyFunction{newNode}
						
						\If{isSymLink}{
							
							linkNode.nextNode $\leftarrow$ newNode;
							
						}
						\ElseIf{hasSymLink}{
							
							newNode.nextNode $\leftarrow$ linkNode;
							
						}
						
						cmdLinkedList.append(newNode);
						
					}
					
				}
				
			}
			
		}
		
	\end{algorithm}
\end{minipage}

\subsubsection{rootfs Pruning}\label{rp}

After obtaining the file information list and the command linked list, we need to prune the fat image's rootfs according to the command token list generated in step 1. First, we match each token in the CTL with the node names in the CLL. If a token $t_{i}$ from CTL matches a node $n_{j}$ in CLL, we designate this node as the starting node, extract the file paths from all subsequent nodes, and store them in the retain list. Additionally, for each path, we use the \texttt{ldd} command to obtain its dependent dynamic link libraries and add them to the retain list. Finally, we add all files from the file information list, excluding those with paths containing \texttt{/bin/}, \texttt{/sbin/}, or \texttt{/lib/}, to the retain list. The retain list is then used to construct the slim \texttt{rootfs.tar} file.

In this section, we address the issue that the tokens in the CTL, as mentioned in Sect. \ref{cte}, do not necessarily consist entirely of command tokens by matching them with the CLL.

\subsection{Slim Image Building}

Based on the static inspection result of the fat image from step 2, we first configure the metadata for the slim image, including the image name, exposed ports, image architecture, etc. Then we use the slim \texttt{rootfs.tar} file to set the layers of the slim image. At this point, $\delta$-SCALPEL generates the corresponding slim image from the fat image.

\section{Evaluation}

We evaluate $\delta$-SCALPEL by answer the following three research questions (RQs):

\begin{itemize}
	\item \textbf{\textit{RQ1}}: What is the effectiveness of $\delta$-SPELCAL?
	
	\item \textbf{\textit{RQ2}}: What is the efficiency of $\delta$-SPELCAL?
	
	\item \textbf{\textit{RQ3}}: What is the significance of image slimming?
\end{itemize}

\subsection{Experiment Setup}

\subsubsection{Dataset} 

To evaluate the $\delta$-SCALPEL model proposed in this paper, we randomly select 20 NPM projects from the top 1000 most depended-upon packages listed on GitHub\footnote{\url{https://gist.github.com/anvaka/8e8fa57c7ee1350e3491}}. For each project, we download its source code from the address specified in the NPM repository and include it in a Dockerfile to create an image. We use Docker Hub's official images, \texttt{node:current-slim} and \texttt{node:current}, to create images for each project. Finally, we create a dataset containing 80 images for model evaluation. In order to verify whether the project runs correctly in the container generated from the slimmed image, we execute the project's test suites using the \texttt{npm test} command and assess its status based on the test suites' results: if the test suites' results before and after slimming are consistent, the image slimming is considered successful; otherwise, it is deemed a failure.

\subsubsection{Implementation} 

All experiments are conducted on a host with 16 CPU cores and 32 GB of memory. We use Go as the development language for $\delta$-SCALPEL, perform static analysis of Node.js code with CodeQL, and utilize the go-dockerclient\footnote{\url{https://github.com/fsouza/go-dockerclient}} package as the client for the Docker remote API.

\subsection{RQ1: Effectiveness}

In a Dockerfile, users can use \texttt{CMD} or \texttt{ENTRYPOINT} to specify the command or script to run when starting a container \cite{dockerfileConcepts}. These two commands are optional. As a result, when building an image with a Dockerfile, users can choose to explicitly specify the container's entry point or leave it unspecified. This section evaluates the effect of $\delta$-SCALPEL. We select Slim \cite{slimtoolkit2024}, a popular and well-established image slimming tool, as the baseline, comparing the effects of $\delta$-SCALPEL and Slim in two scenarios: specifying the entry point and not specifying the entry point.

\subsubsection{RQ1.1: Specify the entry point} 

We specify the entry point of the image for each project in its Dockerfile as \texttt{CMD ["npm", "test"]}, then build project images based on the base images \texttt{node:current-slim} and \texttt{node:current}, respectively. For the generated project images, we use $\delta$-SCALPEL and Slim to perform image slimming. We then start the slimmed images using the \texttt{docker run} command and observe the container's running state.

The evaluation results are shown in Tab. \ref{rq1.1}. For images based on the \texttt{node:current-slim}, using $\delta$-SCALPEL can reduce their size by over 10\% (except for the strip-ansi project image, which sees a reduction of 9.7\%), with the maximum reduction reaching 26.5\%. In contrast, Slim maintains a slimming rate above 50\% for these images (except for the node-portfinder image, which is at 45.1\%). When the base image is \texttt{node:current}, $\delta$-SCALPEL achieves a slimming rate of over 46\%, while Slim exceeds 89\%. However, among the images slimmed by Slim, only the nodejs-websocket and prompt projects can run correctly in the generated container. In contrast, all these projects function normally in the containers created from images slimmed by $\delta$-SCALPEL.

\begin{table*}[!h]\tiny
	\centering
	\caption{Comparison of the image slimming effect between $\delta$-SPELCAL and Slim when the entry point is explicitly specified. $\bullet$ indicates that the slimmed image runs normally, while $\circ$ indicates that it fails to run normally. The best slimming ratio of $\delta$-SCALPEL are \textbf{BOLDED}.}\label{rq1.1}
	\begin{tabular}{|c|c|cccccc|}
		\hline
		\multirow{3}{*}{\textbf{Project}}  & \multirow{3}{*}{\textbf{Model}} & \multicolumn{6}{c|}{\textbf{Basic Image (With Entry Point)}}                                                                                                                                                                                                                                              \\ \cline{3-8} 
		&                                 & \multicolumn{3}{c|}{\textbf{node:current-slim}}                                                                                                               & \multicolumn{3}{c|}{\textbf{node:current}}                                                                                                \\ \cline{3-8} 
		&                                 & \multicolumn{1}{c|}{\textbf{\makecell{Original\\Size}}} & \multicolumn{1}{c|}{\textbf{\makecell{Size After Slimming\\/Slimming Ratio}}} & \multicolumn{1}{c|}{\textbf{\makecell{Running\\Status}}} & \multicolumn{1}{c|}{\textbf{\makecell{Original\\Size}}}  & \multicolumn{1}{c|}{\textbf{\makecell{Size After Slimming\\/Slimming Ratio}}} & \textbf{\makecell{Running\\Status}} \\ \hline
		\multirow{2}{*}{semver}            & $\delta$-SPELCAL                & \multicolumn{1}{c|}{\multirow{2}{*}{551MB}} & \multicolumn{1}{c|}{492MB/10.7\%}                                & \multicolumn{1}{c|}{$\bullet$}               & \multicolumn{1}{c|}{\multirow{2}{*}{1.45GB}} & \multicolumn{1}{c|}{769MB/48.2\%}                                & $\bullet$               \\ \cline{2-2} \cline{4-5} \cline{7-8} 
		& Slim                            & \multicolumn{1}{c|}{}                       & \multicolumn{1}{c|}{126MB/77.1\%}                                & \multicolumn{1}{c|}{$\circ$}                 & \multicolumn{1}{c|}{}                        & \multicolumn{1}{c|}{124MB/91.6\%}                                & $\circ$                 \\ \hline
		\multirow{2}{*}{chalk}             & $\delta$-SPELCAL                & \multicolumn{1}{c|}{\multirow{2}{*}{555MB}} & \multicolumn{1}{c|}{496MB/10.6\%}                                & \multicolumn{1}{c|}{$\bullet$}               & \multicolumn{1}{c|}{\multirow{2}{*}{1.45GB}} & \multicolumn{1}{c|}{772MB/48\%}                                  & $\bullet$               \\ \cline{2-2} \cline{4-5} \cline{7-8} 
		& Slim                            & \multicolumn{1}{c|}{}                       & \multicolumn{1}{c|}{125MB/77.5\%}                                & \multicolumn{1}{c|}{$\circ$}                 & \multicolumn{1}{c|}{}                        & \multicolumn{1}{c|}{125MB/91.6\%}                                & $\circ$                 \\ \hline
		\multirow{2}{*}{nodejs-websocket}  & $\delta$-SPELCAL                & \multicolumn{1}{c|}{\multirow{2}{*}{227MB}} & \multicolumn{1}{c|}{168MB/26\%}                                  & \multicolumn{1}{c|}{$\bullet$}               & \multicolumn{1}{c|}{\multirow{2}{*}{1.12GB}} & \multicolumn{1}{c|}{445MB/61.2\%}                                & $\bullet$               \\ \cline{2-2} \cline{4-5} \cline{7-8} 
		& Slim                            & \multicolumn{1}{c|}{}                       & \multicolumn{1}{c|}{124MB/45.4\%}                                & \multicolumn{1}{c|}{$\bullet$}               & \multicolumn{1}{c|}{}                        & \multicolumn{1}{c|}{124MB/89.2\%}                                & $\bullet$               \\ \hline
		\multirow{2}{*}{lru-cache}         & $\delta$-SPELCAL                & \multicolumn{1}{c|}{\multirow{2}{*}{372MB}} & \multicolumn{1}{c|}{321MB/13.7\%}                                & \multicolumn{1}{c|}{$\bullet$}               & \multicolumn{1}{c|}{\multirow{2}{*}{1.27GB}} & \multicolumn{1}{c|}{598MB/54\%}                                  & $\bullet$               \\ \cline{2-2} \cline{4-5} \cline{7-8} 
		& Slim                            & \multicolumn{1}{c|}{}                       & \multicolumn{1}{c|}{133MB/64.2\%}                                & \multicolumn{1}{c|}{$\circ$}                 & \multicolumn{1}{c|}{}                        & \multicolumn{1}{c|}{124MB/90.4\%}                                & $\circ$                 \\ \hline
		\multirow{2}{*}{minimatch}         & $\delta$-SPELCAL                & \multicolumn{1}{c|}{\multirow{2}{*}{332MB}} & \multicolumn{1}{c|}{272MB/18.1\%}                                & \multicolumn{1}{c|}{$\bullet$}               & \multicolumn{1}{c|}{\multirow{2}{*}{1.23GB}} & \multicolumn{1}{c|}{549MB/56.4\%}                                & $\bullet$               \\ \cline{2-2} \cline{4-5} \cline{7-8} 
		& Slim                            & \multicolumn{1}{c|}{}                       & \multicolumn{1}{c|}{123MB/63\%}                                  & \multicolumn{1}{c|}{$\circ$}                 & \multicolumn{1}{c|}{}                        & \multicolumn{1}{c|}{133MB/89.4\%}                                & $\circ$                 \\ \hline
		\multirow{2}{*}{strip-ansi}        & $\delta$-SPELCAL                & \multicolumn{1}{c|}{\multirow{2}{*}{607MB}} & \multicolumn{1}{c|}{548MB/9.7\%}                                 & \multicolumn{1}{c|}{$\bullet$}               & \multicolumn{1}{c|}{\multirow{2}{*}{1.50GB}}  & \multicolumn{1}{c|}{825MB/46.3\%}                                & $\bullet$               \\ \cline{2-2} \cline{4-5} \cline{7-8} 
		& Slim                            & \multicolumn{1}{c|}{}                       & \multicolumn{1}{c|}{124MB/79.6\%}                                & \multicolumn{1}{c|}{$\circ$}                 & \multicolumn{1}{c|}{}                        & \multicolumn{1}{c|}{124MB/91.9\%}                                & $\circ$                 \\ \hline
		\multirow{2}{*}{node-glob}         & $\delta$-SPELCAL                & \multicolumn{1}{c|}{\multirow{2}{*}{335MB}} & \multicolumn{1}{c|}{275MB/17.9\%}                                & \multicolumn{1}{c|}{$\bullet$}               & \multicolumn{1}{c|}{\multirow{2}{*}{1.23GB}} & \multicolumn{1}{c|}{552MB/56.2\%}                                & $\bullet$               \\ \cline{2-2} \cline{4-5} \cline{7-8} 
		& Slim                            & \multicolumn{1}{c|}{}                       & \multicolumn{1}{c|}{133MB/60.3\%}                                & \multicolumn{1}{c|}{$\circ$}                 & \multicolumn{1}{c|}{}                        & \multicolumn{1}{c|}{133MB/89.4\%}                                & $\circ$                 \\ \hline
		\multirow{2}{*}{commander.js}      & $\delta$-SPELCAL                & \multicolumn{1}{c|}{\multirow{2}{*}{389MB}} & \multicolumn{1}{c|}{328MB/15.7\%}                                & \multicolumn{1}{c|}{$\bullet$}               & \multicolumn{1}{c|}{\multirow{2}{*}{1.29GB}} & \multicolumn{1}{c|}{605MB/54.2\%}                                & $\bullet$               \\ \cline{2-2} \cline{4-5} \cline{7-8} 
		& Slim                            & \multicolumn{1}{c|}{}                       & \multicolumn{1}{c|}{124MB/68.1\%}                                & \multicolumn{1}{c|}{$\circ$}                 & \multicolumn{1}{c|}{}                        & \multicolumn{1}{c|}{125MB/90.5\%}                                & $\circ$                 \\ \hline
		\multirow{2}{*}{yallist}           & $\delta$-SPELCAL                & \multicolumn{1}{c|}{\multirow{2}{*}{384MB}} & \multicolumn{1}{c|}{324MB/15.6\%}                                & \multicolumn{1}{c|}{$\bullet$}               & \multicolumn{1}{c|}{\multirow{2}{*}{1.28GB}} & \multicolumn{1}{c|}{601MB/54.1\%}                                & $\bullet$               \\ \cline{2-2} \cline{4-5} \cline{7-8} 
		& Slim                            & \multicolumn{1}{c|}{}                       & \multicolumn{1}{c|}{123MB/68\%}                                  & \multicolumn{1}{c|}{$\circ$}                 & \multicolumn{1}{c|}{}                        & \multicolumn{1}{c|}{133MB/89.9\%}                                & $\circ$                 \\ \hline
		\multirow{2}{*}{estraverse}        & $\delta$-SPELCAL                & \multicolumn{1}{c|}{\multirow{2}{*}{276MB}} & \multicolumn{1}{c|}{216MB/21.7\%}                                & \multicolumn{1}{c|}{$\bullet$}               & \multicolumn{1}{c|}{\multirow{2}{*}{1.17GB}} & \multicolumn{1}{c|}{493MB/58.9\%}                                & $\bullet$               \\ \cline{2-2} \cline{4-5} \cline{7-8} 
		& Slim                            & \multicolumn{1}{c|}{}                       & \multicolumn{1}{c|}{123MB/55.4\%}                                & \multicolumn{1}{c|}{$\circ$}                 & \multicolumn{1}{c|}{}                        & \multicolumn{1}{c|}{124MB/89.7\%}                                & $\circ$                 \\ \hline
		\multirow{2}{*}{deepmerge}         & $\delta$-SPELCAL                & \multicolumn{1}{c|}{\multirow{2}{*}{263MB}} & \multicolumn{1}{c|}{202MB/23.2\%}                                & \multicolumn{1}{c|}{$\bullet$}               & \multicolumn{1}{c|}{\multirow{2}{*}{1.16GB}} & \multicolumn{1}{c|}{479MB/59.7\%}                                & $\bullet$               \\ \cline{2-2} \cline{4-5} \cline{7-8} 
		& Slim                            & \multicolumn{1}{c|}{}                       & \multicolumn{1}{c|}{125MB/52.5\%}                                & \multicolumn{1}{c|}{$\circ$}                 & \multicolumn{1}{c|}{}                        & \multicolumn{1}{c|}{126MB/89.4\%}                                & $\circ$                 \\ \hline
		\multirow{2}{*}{node-fs-extra}     & $\delta$-SPELCAL                & \multicolumn{1}{c|}{\multirow{2}{*}{315MB}} & \multicolumn{1}{c|}{254MB/19.4\%}                                & \multicolumn{1}{c|}{$\bullet$}               & \multicolumn{1}{c|}{\multirow{2}{*}{1.21GB}} & \multicolumn{1}{c|}{531MB/57.1\%}                                & $\bullet$               \\ \cline{2-2} \cline{4-5} \cline{7-8} 
		& Slim                            & \multicolumn{1}{c|}{}                       & \multicolumn{1}{c|}{135MB/57.1\%}                                & \multicolumn{1}{c|}{$\circ$}                 & \multicolumn{1}{c|}{}                        & \multicolumn{1}{c|}{124MB/90\%}                                  & $\circ$                 \\ \hline
		\multirow{2}{*}{node-jsonwebtoken} & $\delta$-SPELCAL                & \multicolumn{1}{c|}{\multirow{2}{*}{313MB}} & \multicolumn{1}{c|}{252MB/19.5\%}                                & \multicolumn{1}{c|}{$\bullet$}               & \multicolumn{1}{c|}{\multirow{2}{*}{1.21GB}} & \multicolumn{1}{c|}{529MB/57.3\%}                                & $\bullet$               \\ \cline{2-2} \cline{4-5} \cline{7-8} 
		& Slim                            & \multicolumn{1}{c|}{}                       & \multicolumn{1}{c|}{123MB/60.7\%}                                & \multicolumn{1}{c|}{$\circ$}                 & \multicolumn{1}{c|}{}                        & \multicolumn{1}{c|}{131MB/89.4\%}                                & $\circ$                 \\ \hline
		\multirow{2}{*}{node-which}        & $\delta$-SPELCAL                & \multicolumn{1}{c|}{\multirow{2}{*}{553MB}} & \multicolumn{1}{c|}{492MB/11\%}                                  & \multicolumn{1}{c|}{$\bullet$}               & \multicolumn{1}{c|}{\multirow{2}{*}{1.45GB}} & \multicolumn{1}{c|}{769MB/48.2\%}                                & $\bullet$               \\ \cline{2-2} \cline{4-5} \cline{7-8} 
		& Slim                            & \multicolumn{1}{c|}{}                       & \multicolumn{1}{c|}{125MB/77.4\%}                                & \multicolumn{1}{c|}{$\circ$}                 & \multicolumn{1}{c|}{}                        & \multicolumn{1}{c|}{126MB/91.5\%}                                & $\circ$                 \\ \hline
		\multirow{2}{*}{prompt}            & $\delta$-SPELCAL                & \multicolumn{1}{c|}{\multirow{2}{*}{253MB}} & \multicolumn{1}{c|}{192MB/24.1\%}                                & \multicolumn{1}{c|}{$\bullet$}               & \multicolumn{1}{c|}{\multirow{2}{*}{1.15GB}} & \multicolumn{1}{c|}{470MB/60.1\%}                                & $\bullet$               \\ \cline{2-2} \cline{4-5} \cline{7-8} 
		& Slim                            & \multicolumn{1}{c|}{}                       & \multicolumn{1}{c|}{124MB/51\%}                                  & \multicolumn{1}{c|}{$\bullet$}               & \multicolumn{1}{c|}{}                        & \multicolumn{1}{c|}{124MB/89.5\%}                                & $\bullet$               \\ \hline
		\multirow{2}{*}{shelljs}           & $\delta$-SPELCAL                & \multicolumn{1}{c|}{\multirow{2}{*}{345MB}} & \multicolumn{1}{c|}{284MB/17.7\%}                                & \multicolumn{1}{c|}{$\bullet$}               & \multicolumn{1}{c|}{\multirow{2}{*}{1.29GB}} & \multicolumn{1}{c|}{613MB/53.6\%}                                & $\bullet$               \\ \cline{2-2} \cline{4-5} \cline{7-8} 
		& Slim                            & \multicolumn{1}{c|}{}                       & \multicolumn{1}{c|}{127MB/63.2\%}                                & \multicolumn{1}{c|}{$\circ$}                 & \multicolumn{1}{c|}{}                        & \multicolumn{1}{c|}{126MB/90.5\%}                                & $\circ$                 \\ \hline
		\multirow{2}{*}{winston}           & $\delta$-SPELCAL                & \multicolumn{1}{c|}{\multirow{2}{*}{292MB}} & \multicolumn{1}{c|}{230MB/21.2\%}                                & \multicolumn{1}{c|}{$\bullet$}               & \multicolumn{1}{c|}{\multirow{2}{*}{1.19GB}} & \multicolumn{1}{c|}{507MB/58.4\%}                                & $\bullet$               \\ \cline{2-2} \cline{4-5} \cline{7-8} 
		& Slim                            & \multicolumn{1}{c|}{}                       & \multicolumn{1}{c|}{125MB/57.2\%}                                & \multicolumn{1}{c|}{$\circ$}                 & \multicolumn{1}{c|}{}                        & \multicolumn{1}{c|}{125MB/89.7\%}                                & $\circ$                 \\ \hline
		\multirow{2}{*}{ws}                & $\delta$-SPELCAL                & \multicolumn{1}{c|}{\multirow{2}{*}{307MB}} & \multicolumn{1}{c|}{246MB/19.9\%}                                & \multicolumn{1}{c|}{$\bullet$}               & \multicolumn{1}{c|}{\multirow{2}{*}{1.2GB}}  & \multicolumn{1}{c|}{523MB/57.4\%}                                & $\bullet$               \\ \cline{2-2} \cline{4-5} \cline{7-8} 
		& Slim                            & \multicolumn{1}{c|}{}                       & \multicolumn{1}{c|}{124MB/59.6\%}                                & \multicolumn{1}{c|}{$\circ$}                 & \multicolumn{1}{c|}{}                        & \multicolumn{1}{c|}{126MB/89.7\%}                                & $\circ$                 \\ \hline
		\multirow{2}{*}{minimist}          & $\delta$-SPELCAL                & \multicolumn{1}{c|}{\multirow{2}{*}{303MB}} & \multicolumn{1}{c|}{242MB/20.1\%}                                & \multicolumn{1}{c|}{$\bullet$}               & \multicolumn{1}{c|}{\multirow{2}{*}{1.2GB}}  & \multicolumn{1}{c|}{519MB/57.8\%}                                & $\bullet$               \\ \cline{2-2} \cline{4-5} \cline{7-8} 
		& Slim                            & \multicolumn{1}{c|}{}                       & \multicolumn{1}{c|}{124MB/59.1\%}                                & \multicolumn{1}{c|}{$\circ$}                 & \multicolumn{1}{c|}{}                        & \multicolumn{1}{c|}{124MB/89.9\%}                                & $\circ$                 \\ \hline
		\multirow{2}{*}{node-portfinder}   & $\delta$-SPELCAL                & \multicolumn{1}{c|}{\multirow{2}{*}{226MB}} & \multicolumn{1}{c|}{\textbf{166MB/26.5}\%}                                & \multicolumn{1}{c|}{$\bullet$}               & \multicolumn{1}{c|}{\multirow{2}{*}{1.12GB}} & \multicolumn{1}{c|}{\textbf{443MB/61.4}\%}                                & $\bullet$               \\ \cline{2-2} \cline{4-5} \cline{7-8} 
		& Slim                            & \multicolumn{1}{c|}{}                       & \multicolumn{1}{c|}{124MB/45.1\%}                                & \multicolumn{1}{c|}{$\circ$}                 & \multicolumn{1}{c|}{}                        & \multicolumn{1}{c|}{124MB/89.2\%}                                & $\circ$                 \\ \hline
	\end{tabular}
\end{table*}

\subsubsection{RQ1.2: Do not specify the entry point}

We use the same method as in RQ1.1 to build images for the 20 NPM projects, based on both \texttt{node:current-slim} and \texttt{node:current}. However, in the Dockerfile, we remove the entry point. Similarly, we use $\delta$-SCALPEL and Slim to slim these images. Afterward, we access the containers generated from these slimmed images, manually enter \texttt{npm test} command in the project root directory to run the test suites, and observe the results.

The evaluation results are shown in Tab. \ref{rq1.2}. When the entry point is not specified, $\delta$-SCALPEL achieves the same slimming effect as when the entry point is specified. All NPM projects in the containers generated from these slimmed images run normally. For the images slimmed by Slim, although the size is reduced by more than 45\% compared to the original (with a maximum reduction of 92\%), Slim fails to capture the environment dependencies required for the project code to run due to the absence of an explicitly specified entry point. Additionally, we find that for images slimmed by Slim, whether the base image is \texttt{node:current-slim} or \texttt{node:current}, the size of the slimmed image is 123MB. This indicates that Slim retains only the base runtime environment for the container when the entry point is not specified.

\begin{table*}[!h]\tiny
	\centering
	\caption{Comparison of the image slimming effect between $\delta$-SPELCAL and Slim when the entry point is not explicitly specified. $\bullet$ indicates that the slimmed image runs normally, while $\circ$ indicates that it fails to run normally. The best slimming ratio of $\delta$-SCALPEL are \textbf{BOLDED}.}\label{rq1.2}
	\begin{tabular}{|c|c|cccccc|}
		\hline
		\multirow{3}{*}{\textbf{Project}}  & \multirow{3}{*}{\textbf{Model}} & \multicolumn{6}{c|}{\textbf{Basic Image (Without Entry Point)}}                                                                                                                                                                                                                                           \\ \cline{3-8} 
		&                                 & \multicolumn{3}{c|}{\textbf{node:current-slim}}                                                                                                               & \multicolumn{3}{c|}{\textbf{node:current}}                                                                                                \\ \cline{3-8} 
		&                                 & \multicolumn{1}{c|}{\textbf{\makecell{Original\\Size}}} & \multicolumn{1}{c|}{\textbf{\makecell{Size After Slimming\\/Slimming Ratio}}} & \multicolumn{1}{c|}{\textbf{\makecell{Running\\Status}}} & \multicolumn{1}{c|}{\textbf{\makecell{Original\\Size}}}  & \multicolumn{1}{c|}{\textbf{\makecell{Size After Slimming\\/Slimming Ratio}}} & \textbf{\makecell{Running\\Status}} \\ \hline
		\multirow{2}{*}{semver}            & $\delta$-SPELCAL                & \multicolumn{1}{c|}{\multirow{2}{*}{551MB}} & \multicolumn{1}{c|}{492MB/10.7\%}                                & \multicolumn{1}{c|}{$\bullet$}               & \multicolumn{1}{c|}{\multirow{2}{*}{1.45GB}} & \multicolumn{1}{c|}{769MB/48.2\%}                                & $\bullet$               \\ \cline{2-2} \cline{4-5} \cline{7-8} 
		& Slim                            & \multicolumn{1}{c|}{}                       & \multicolumn{1}{c|}{123MB/77.7\%}                                & \multicolumn{1}{c|}{$\circ$}                 & \multicolumn{1}{c|}{}                        & \multicolumn{1}{c|}{123MB/91.7\%}                                & $\circ$                 \\ \hline
		\multirow{2}{*}{chalk}             & $\delta$-SPELCAL                & \multicolumn{1}{c|}{\multirow{2}{*}{555MB}} & \multicolumn{1}{c|}{496MB/10.6\%}                                & \multicolumn{1}{c|}{$\bullet$}               & \multicolumn{1}{c|}{\multirow{2}{*}{1.45GB}} & \multicolumn{1}{c|}{772MB/48\%}                                  & $\bullet$               \\ \cline{2-2} \cline{4-5} \cline{7-8} 
		& Slim                            & \multicolumn{1}{c|}{}                       & \multicolumn{1}{c|}{123MB/77.9\%}                                & \multicolumn{1}{c|}{$\circ$}                 & \multicolumn{1}{c|}{}                        & \multicolumn{1}{c|}{123MB/91.7\%}                                & $\circ$                 \\ \hline
		\multirow{2}{*}{nodejs-websocket}  & $\delta$-SPELCAL                & \multicolumn{1}{c|}{\multirow{2}{*}{227MB}} & \multicolumn{1}{c|}{168MB/26\%}                                  & \multicolumn{1}{c|}{$\bullet$}               & \multicolumn{1}{c|}{\multirow{2}{*}{1.12GB}} & \multicolumn{1}{c|}{445MB/61.2\%}                                & $\bullet$               \\ \cline{2-2} \cline{4-5} \cline{7-8} 
		& Slim                            & \multicolumn{1}{c|}{}                       & \multicolumn{1}{c|}{123MB/45.8\%}                                & \multicolumn{1}{c|}{$\circ$}                 & \multicolumn{1}{c|}{}                        & \multicolumn{1}{c|}{123MB/89.3\%}                                & $\circ$                 \\ \hline
		\multirow{2}{*}{lru-cache}         & $\delta$-SPELCAL                & \multicolumn{1}{c|}{\multirow{2}{*}{372MB}} & \multicolumn{1}{c|}{321MB/13.7\%}                                & \multicolumn{1}{c|}{$\bullet$}               & \multicolumn{1}{c|}{\multirow{2}{*}{1.27GB}} & \multicolumn{1}{c|}{598MB/54\%}                                  & $\bullet$               \\ \cline{2-2} \cline{4-5} \cline{7-8} 
		& Slim                            & \multicolumn{1}{c|}{}                       & \multicolumn{1}{c|}{123MB/66.9\%}                                & \multicolumn{1}{c|}{$\circ$}                 & \multicolumn{1}{c|}{}                        & \multicolumn{1}{c|}{123MB/90.5\%}                                & $\circ$                 \\ \hline
		\multirow{2}{*}{minimatch}         & $\delta$-SPELCAL                & \multicolumn{1}{c|}{\multirow{2}{*}{332MB}} & \multicolumn{1}{c|}{272MB/18.1\%}                                & \multicolumn{1}{c|}{$\bullet$}               & \multicolumn{1}{c|}{\multirow{2}{*}{1.23GB}} & \multicolumn{1}{c|}{549MB/56.4\%}                                & $\bullet$               \\ \cline{2-2} \cline{4-5} \cline{7-8} 
		& Slim                            & \multicolumn{1}{c|}{}                       & \multicolumn{1}{c|}{123MB/63\%}                                  & \multicolumn{1}{c|}{$\circ$}                 & \multicolumn{1}{c|}{}                        & \multicolumn{1}{c|}{123MB/90.2\%}                                & $\circ$                 \\ \hline
		\multirow{2}{*}{strip-ansi}        & $\delta$-SPELCAL                & \multicolumn{1}{c|}{\multirow{2}{*}{607MB}} & \multicolumn{1}{c|}{548MB/9.7\%}                                 & \multicolumn{1}{c|}{$\bullet$}               & \multicolumn{1}{c|}{\multirow{2}{*}{1.5GB}}  & \multicolumn{1}{c|}{825MB/46.3\%}                                & $\bullet$               \\ \cline{2-2} \cline{4-5} \cline{7-8} 
		& Slim                            & \multicolumn{1}{c|}{}                       & \multicolumn{1}{c|}{123MB/79.7\%}                                & \multicolumn{1}{c|}{$\circ$}                 & \multicolumn{1}{c|}{}                        & \multicolumn{1}{c|}{123MB/92\%}                                  & $\circ$                 \\ \hline
		\multirow{2}{*}{node-glob}         & $\delta$-SPELCAL                & \multicolumn{1}{c|}{\multirow{2}{*}{335MB}} & \multicolumn{1}{c|}{275MB/17.9\%}                                & \multicolumn{1}{c|}{$\bullet$}               & \multicolumn{1}{c|}{\multirow{2}{*}{1.23GB}} & \multicolumn{1}{c|}{552MB/56.2\%}                                & $\bullet$               \\ \cline{2-2} \cline{4-5} \cline{7-8} 
		& Slim                            & \multicolumn{1}{c|}{}                       & \multicolumn{1}{c|}{123MB/63.3\%}                                & \multicolumn{1}{c|}{$\circ$}                 & \multicolumn{1}{c|}{}                        & \multicolumn{1}{c|}{123MB/90\%}                                  & $\circ$                 \\ \hline
		\multirow{2}{*}{commander.js}      & $\delta$-SPELCAL                & \multicolumn{1}{c|}{\multirow{2}{*}{389MB}} & \multicolumn{1}{c|}{328MB/15.7\%}                                & \multicolumn{1}{c|}{$\bullet$}               & \multicolumn{1}{c|}{\multirow{2}{*}{1.29GB}} & \multicolumn{1}{c|}{605MB/54.2\%}                                & $\bullet$               \\ \cline{2-2} \cline{4-5} \cline{7-8} 
		& Slim                            & \multicolumn{1}{c|}{}                       & \multicolumn{1}{c|}{123MB/68.4\%}                                & \multicolumn{1}{c|}{$\circ$}                 & \multicolumn{1}{c|}{}                        & \multicolumn{1}{c|}{123MB/90.5\%}                                & $\circ$                 \\ \hline
		\multirow{2}{*}{yallist}           & $\delta$-SPELCAL                & \multicolumn{1}{c|}{\multirow{2}{*}{384MB}} & \multicolumn{1}{c|}{324MB/15.6\%}                                & \multicolumn{1}{c|}{$\bullet$}               & \multicolumn{1}{c|}{\multirow{2}{*}{1.28GB}} & \multicolumn{1}{c|}{601MB/54.1\%}                                & $\bullet$               \\ \cline{2-2} \cline{4-5} \cline{7-8} 
		& Slim                            & \multicolumn{1}{c|}{}                       & \multicolumn{1}{c|}{123MB/68\%}                                  & \multicolumn{1}{c|}{$\circ$}                 & \multicolumn{1}{c|}{}                        & \multicolumn{1}{c|}{123MB/90.6\%}                                & $\circ$                 \\ \hline
		\multirow{2}{*}{estraverse}        & $\delta$-SPELCAL                & \multicolumn{1}{c|}{\multirow{2}{*}{276MB}} & \multicolumn{1}{c|}{216MB/21.7\%}                                & \multicolumn{1}{c|}{$\bullet$}               & \multicolumn{1}{c|}{\multirow{2}{*}{1.17GB}} & \multicolumn{1}{c|}{493MB/58.9\%}                                & $\bullet$               \\ \cline{2-2} \cline{4-5} \cline{7-8} 
		& Slim                            & \multicolumn{1}{c|}{}                       & \multicolumn{1}{c|}{123MB/55.4\%}                                & \multicolumn{1}{c|}{$\circ$}                 & \multicolumn{1}{c|}{}                        & \multicolumn{1}{c|}{123MB/89.7\%}                                & $\circ$                 \\ \hline
		\multirow{2}{*}{deepmerge}         & $\delta$-SPELCAL                & \multicolumn{1}{c|}{\multirow{2}{*}{263MB}} & \multicolumn{1}{c|}{202MB/23.2\%}                                & \multicolumn{1}{c|}{$\bullet$}               & \multicolumn{1}{c|}{\multirow{2}{*}{1.16GB}} & \multicolumn{1}{c|}{479MB/59.7\%}                                & $\bullet$               \\ \cline{2-2} \cline{4-5} \cline{7-8} 
		& Slim                            & \multicolumn{1}{c|}{}                       & \multicolumn{1}{c|}{123MB/53.2\%}                                & \multicolumn{1}{c|}{$\circ$}                 & \multicolumn{1}{c|}{}                        & \multicolumn{1}{c|}{123MB/89.6\%}                                & $\circ$                 \\ \hline
		\multirow{2}{*}{node-fs-extra}     & $\delta$-SPELCAL                & \multicolumn{1}{c|}{\multirow{2}{*}{315MB}} & \multicolumn{1}{c|}{254MB/19.4\%}                                & \multicolumn{1}{c|}{$\bullet$}               & \multicolumn{1}{c|}{\multirow{2}{*}{1.21GB}} & \multicolumn{1}{c|}{531MB/57.1\%}                                & $\bullet$               \\ \cline{2-2} \cline{4-5} \cline{7-8} 
		& Slim                            & \multicolumn{1}{c|}{}                       & \multicolumn{1}{c|}{123MB/61\%}                                  & \multicolumn{1}{c|}{$\circ$}                 & \multicolumn{1}{c|}{}                        & \multicolumn{1}{c|}{123MB/90.1\%}                                & $\circ$                 \\ \hline
		\multirow{2}{*}{node-jsonwebtoken} & $\delta$-SPELCAL                & \multicolumn{1}{c|}{\multirow{2}{*}{313MB}} & \multicolumn{1}{c|}{252MB/19.5\%}                                & \multicolumn{1}{c|}{$\bullet$}               & \multicolumn{1}{c|}{\multirow{2}{*}{1.21GB}} & \multicolumn{1}{c|}{529MB/57.3\%}                                & $\bullet$               \\ \cline{2-2} \cline{4-5} \cline{7-8} 
		& Slim                            & \multicolumn{1}{c|}{}                       & \multicolumn{1}{c|}{123MB/60.7\%}                                & \multicolumn{1}{c|}{$\circ$}                 & \multicolumn{1}{c|}{}                        & \multicolumn{1}{c|}{123MB/90.1\%}                                & $\circ$                 \\ \hline
		\multirow{2}{*}{node-which}        & $\delta$-SPELCAL                & \multicolumn{1}{c|}{\multirow{2}{*}{553MB}} & \multicolumn{1}{c|}{492MB/11\%}                                  & \multicolumn{1}{c|}{$\bullet$}               & \multicolumn{1}{c|}{\multirow{2}{*}{1.45GB}} & \multicolumn{1}{c|}{769MB/48.2\%}                                & $\bullet$               \\ \cline{2-2} \cline{4-5} \cline{7-8} 
		& Slim                            & \multicolumn{1}{c|}{}                       & \multicolumn{1}{c|}{123MB/77.8\%}                                & \multicolumn{1}{c|}{$\circ$}                 & \multicolumn{1}{c|}{}                        & \multicolumn{1}{c|}{123MB/91.7\%}                                & $\circ$                 \\ \hline
		\multirow{2}{*}{prompt}            & $\delta$-SPELCAL                & \multicolumn{1}{c|}{\multirow{2}{*}{253MB}} & \multicolumn{1}{c|}{192MB/24.1\%}                                & \multicolumn{1}{c|}{$\bullet$}               & \multicolumn{1}{c|}{\multirow{2}{*}{1.15GB}} & \multicolumn{1}{c|}{470MB/60.1\%}                                & $\bullet$               \\ \cline{2-2} \cline{4-5} \cline{7-8} 
		& Slim                            & \multicolumn{1}{c|}{}                       & \multicolumn{1}{c|}{123MB/51.4\%}                                & \multicolumn{1}{c|}{$\circ$}                 & \multicolumn{1}{c|}{}                        & \multicolumn{1}{c|}{123MB/89.6\%}                                & $\circ$                 \\ \hline
		\multirow{2}{*}{shelljs}           & $\delta$-SPELCAL                & \multicolumn{1}{c|}{\multirow{2}{*}{345MB}} & \multicolumn{1}{c|}{284MB/17.7\%}                                & \multicolumn{1}{c|}{$\bullet$}               & \multicolumn{1}{c|}{\multirow{2}{*}{1.29GB}} & \multicolumn{1}{c|}{613MB/53.6\%}                                & $\bullet$               \\ \cline{2-2} \cline{4-5} \cline{7-8} 
		& Slim                            & \multicolumn{1}{c|}{}                       & \multicolumn{1}{c|}{123MB/64.3\%}                                & \multicolumn{1}{c|}{$\circ$}                 & \multicolumn{1}{c|}{}                        & \multicolumn{1}{c|}{123MB/90.7\%}                                & $\circ$                 \\ \hline
		\multirow{2}{*}{winston}           & $\delta$-SPELCAL                & \multicolumn{1}{c|}{\multirow{2}{*}{292MB}} & \multicolumn{1}{c|}{230MB/21.2\%}                                & \multicolumn{1}{c|}{$\bullet$}               & \multicolumn{1}{c|}{\multirow{2}{*}{1.19GB}} & \multicolumn{1}{c|}{507MB/58.4\%}                                & $\bullet$               \\ \cline{2-2} \cline{4-5} \cline{7-8} 
		& Slim                            & \multicolumn{1}{c|}{}                       & \multicolumn{1}{c|}{123MB/57.9\%}                                & \multicolumn{1}{c|}{$\circ$}                 & \multicolumn{1}{c|}{}                        & \multicolumn{1}{c|}{123MB/89.9\%}                                & $\circ$                 \\ \hline
		\multirow{2}{*}{ws}                & $\delta$-SPELCAL                & \multicolumn{1}{c|}{\multirow{2}{*}{307MB}} & \multicolumn{1}{c|}{246MB/19.9\%}                                & \multicolumn{1}{c|}{$\bullet$}               & \multicolumn{1}{c|}{\multirow{2}{*}{1.2GB}}  & \multicolumn{1}{c|}{523MB/57.4\%}                                & $\bullet$               \\ \cline{2-2} \cline{4-5} \cline{7-8} 
		& Slim                            & \multicolumn{1}{c|}{}                       & \multicolumn{1}{c|}{123MB/59.9\%}                                & \multicolumn{1}{c|}{$\circ$}                 & \multicolumn{1}{c|}{}                        & \multicolumn{1}{c|}{123MB/90\%}                                  & $\circ$                 \\ \hline
		\multirow{2}{*}{minimist}          & $\delta$-SPELCAL                & \multicolumn{1}{c|}{\multirow{2}{*}{303MB}} & \multicolumn{1}{c|}{242MB/20.1\%}                                & \multicolumn{1}{c|}{$\bullet$}               & \multicolumn{1}{c|}{\multirow{2}{*}{1.2GB}}  & \multicolumn{1}{c|}{519MB/57.8\%}                                & $\bullet$               \\ \cline{2-2} \cline{4-5} \cline{7-8} 
		& Slim                            & \multicolumn{1}{c|}{}                       & \multicolumn{1}{c|}{123MB/59.4\%}                                & \multicolumn{1}{c|}{$\circ$}                 & \multicolumn{1}{c|}{}                        & \multicolumn{1}{c|}{123MB/90\%}                                  & $\circ$                 \\ \hline
		\multirow{2}{*}{node-portfinder}   & $\delta$-SPELCAL                & \multicolumn{1}{c|}{\multirow{2}{*}{226MB}} & \multicolumn{1}{c|}{\textbf{166MB/26.5}\%}                                & \multicolumn{1}{c|}{$\bullet$}               & \multicolumn{1}{c|}{\multirow{2}{*}{1.12GB}} & \multicolumn{1}{c|}{\textbf{443MB/61.4}\%}                                & $\bullet$               \\ \cline{2-2} \cline{4-5} \cline{7-8} 
		& Slim                            & \multicolumn{1}{c|}{}                       & \multicolumn{1}{c|}{123MB/45.6\%}                                & \multicolumn{1}{c|}{$\circ$}                 & \multicolumn{1}{c|}{}                        & \multicolumn{1}{c|}{123MB/89.3\%}                                & $\circ$                 \\ \hline
	\end{tabular}
\end{table*}

\newtcolorbox{mybox}{colframe = red!75!black}
\begin{mybox}
	\textbf{Answer to RQ1:} Regardless of whether the image entry point is specified, $\delta$-SCALPEL ensures that the project code in the container generated by the slimmed image runs normally.
\end{mybox}

\subsection{RQ2: Efficiency}

Based on the $\delta$-SCALPEL's framework, we divide it into four parts: code analyzer preparation (CA pre, we refer to the three modules in step 1 collectively as the code analyzer), code analyzer execution (CA exec), image analyzer preparation and execution (IA pre \& exec), and slim image building (SI building). In the scenario where the image entry point is specified, we evaluate the efficiency of $\delta$-SCALPEL using the base images \texttt{node:current-slim} and \texttt{node:current}, respectively.

\begin{figure*}[!t]
	\centering
	\subfigure[semver]{
		\centering
		\includegraphics[width=0.18\textwidth]{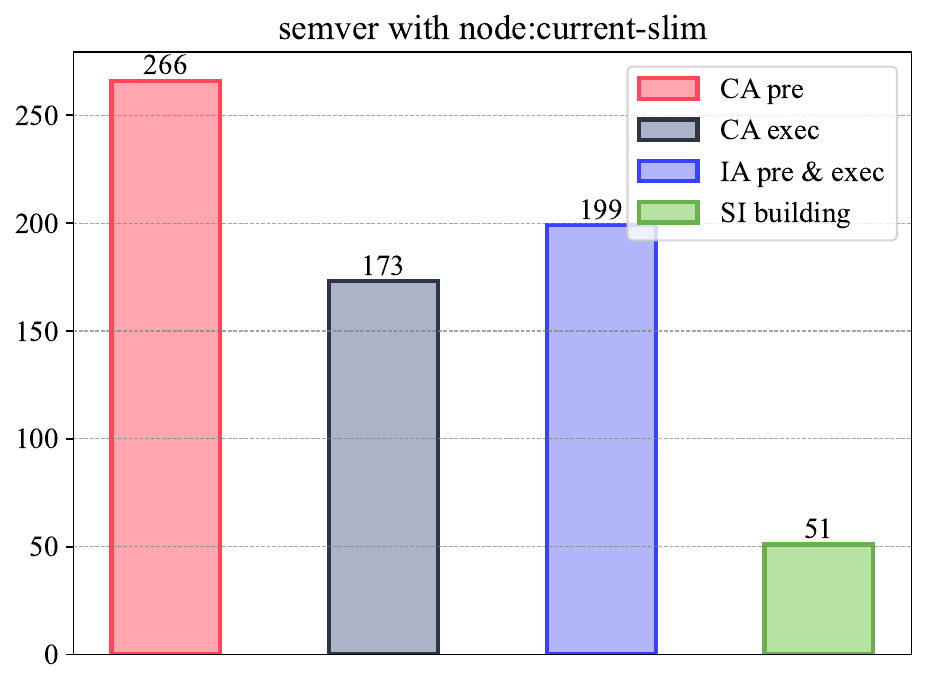}
	}
	\subfigure[chalk]{
		\centering
		\includegraphics[width=0.18\textwidth]{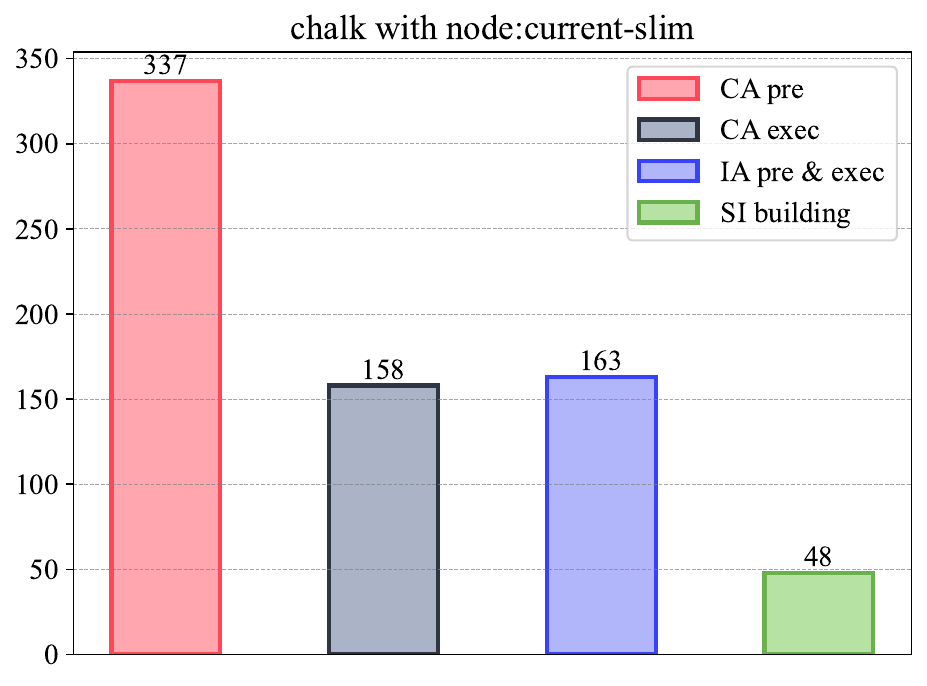}
	}
	\subfigure[nodejs-websocket]{
		\centering
		\includegraphics[width=0.18\textwidth]{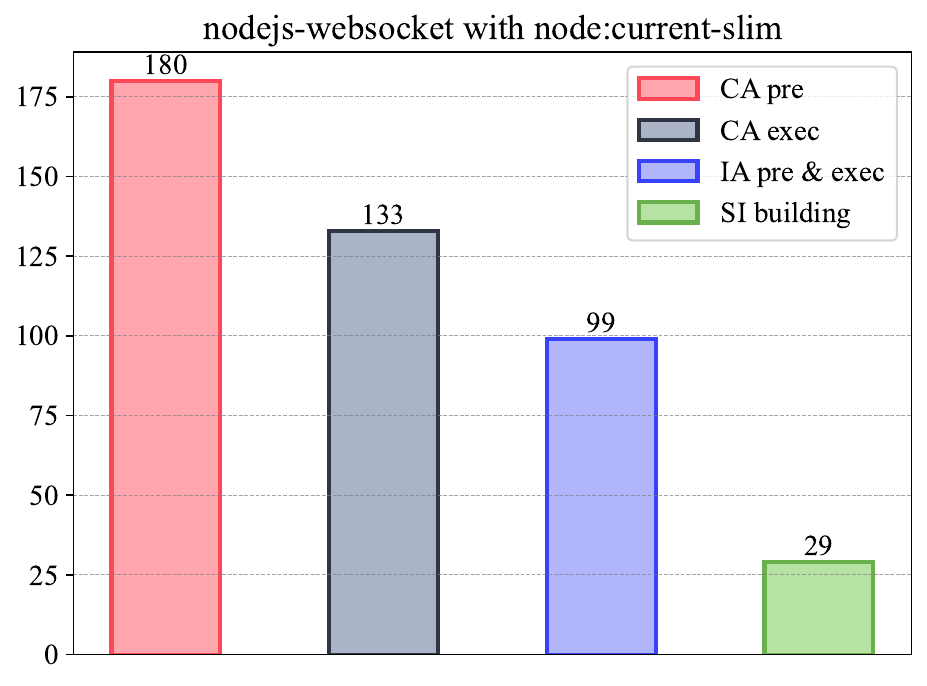}
	}
	\subfigure[lru-cache]{
		\centering
		\includegraphics[width=0.18\textwidth]{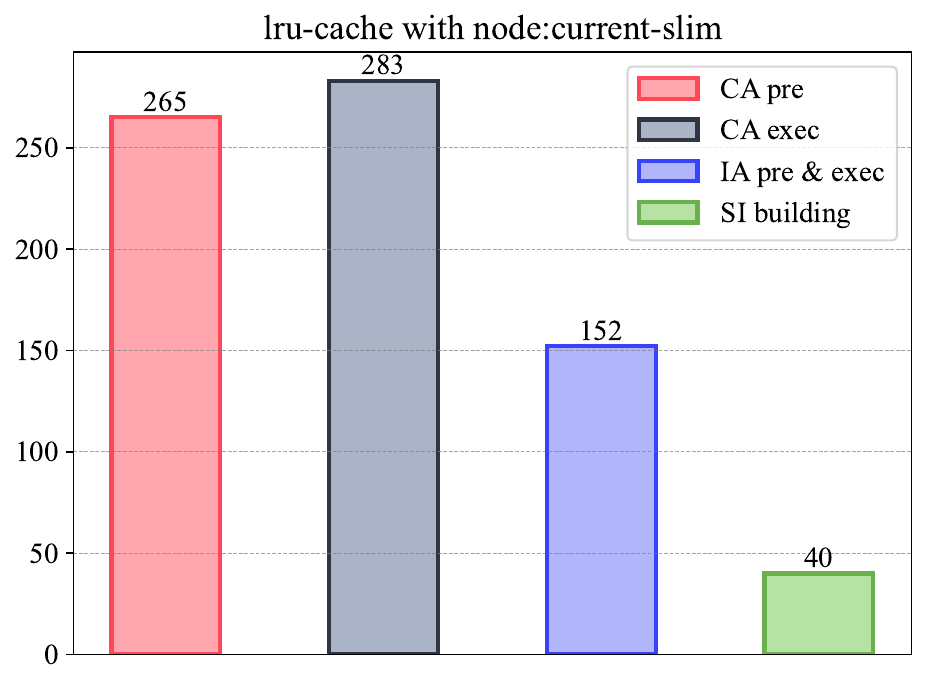}
	}
	\subfigure[minimatch]{
		\centering
		\includegraphics[width=0.18\textwidth]{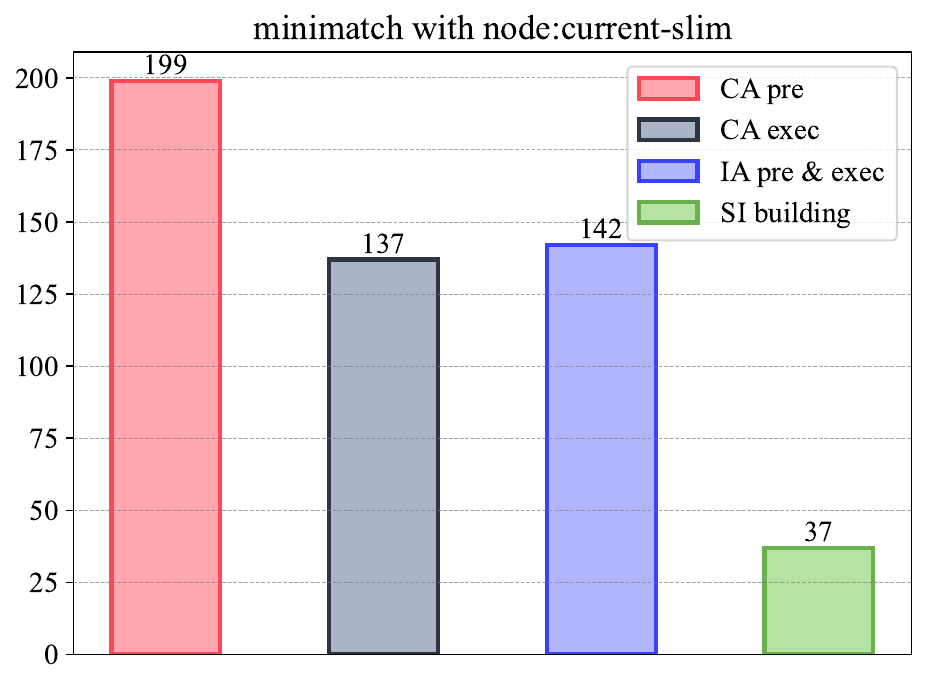}
	}
	
	\subfigure[strip-ansi]{
		\centering
		\includegraphics[width=0.18\textwidth]{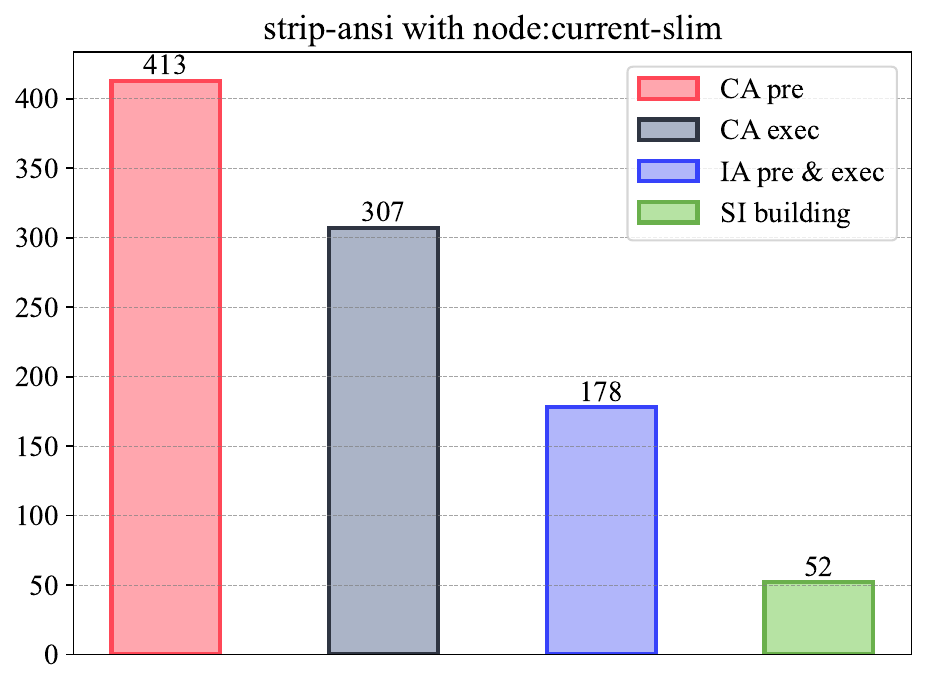}
	}
	\subfigure[node-glob]{
		\centering
		\includegraphics[width=0.18\textwidth]{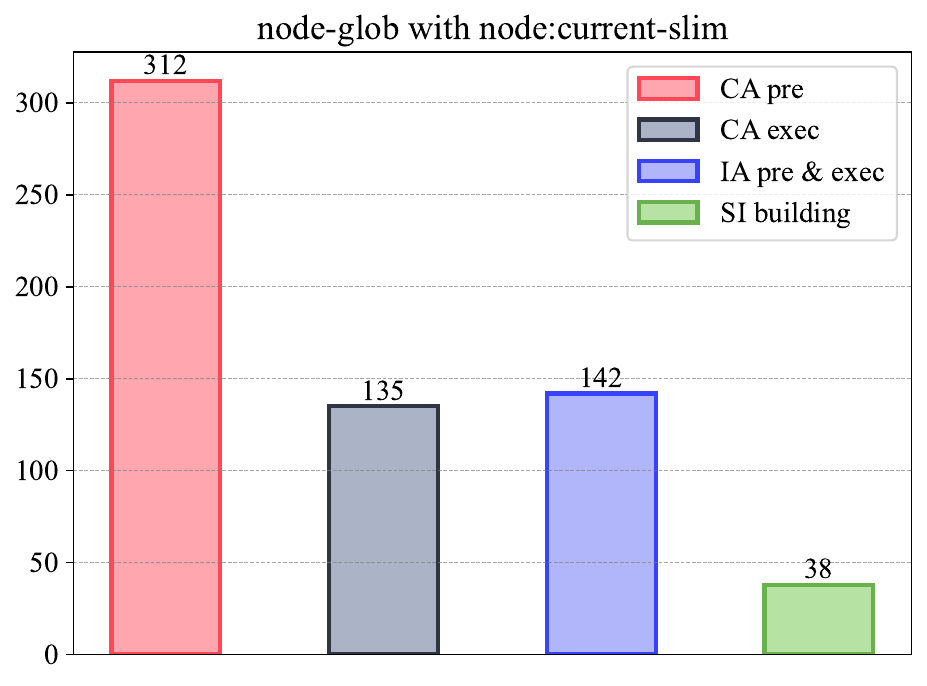}
	}
	\subfigure[commander.js]{
		\centering
		\includegraphics[width=0.18\textwidth]{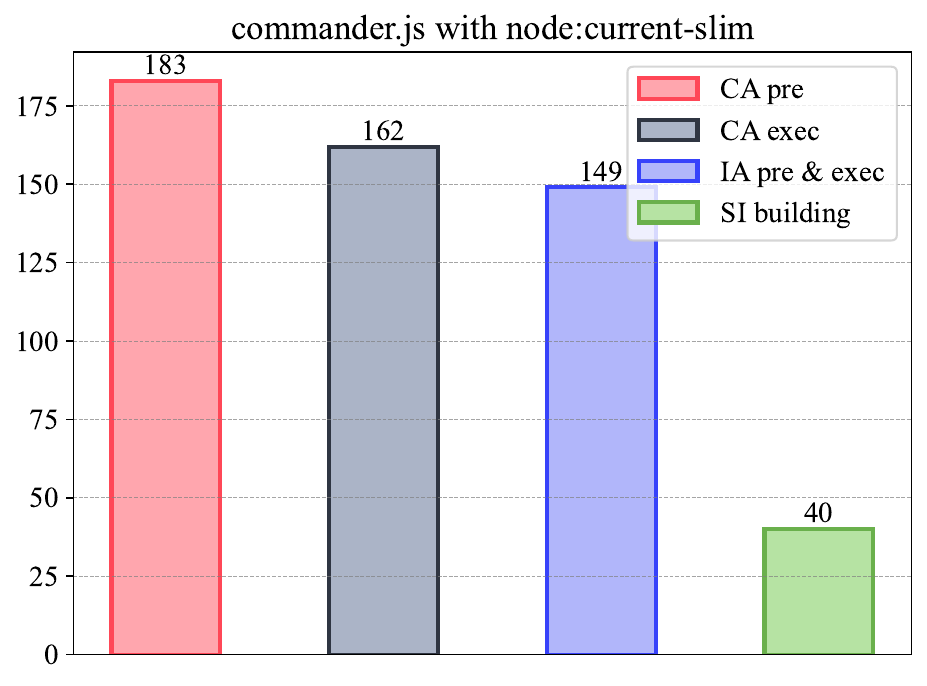}
	}
	\subfigure[yallist]{
		\centering
		\includegraphics[width=0.18\textwidth]{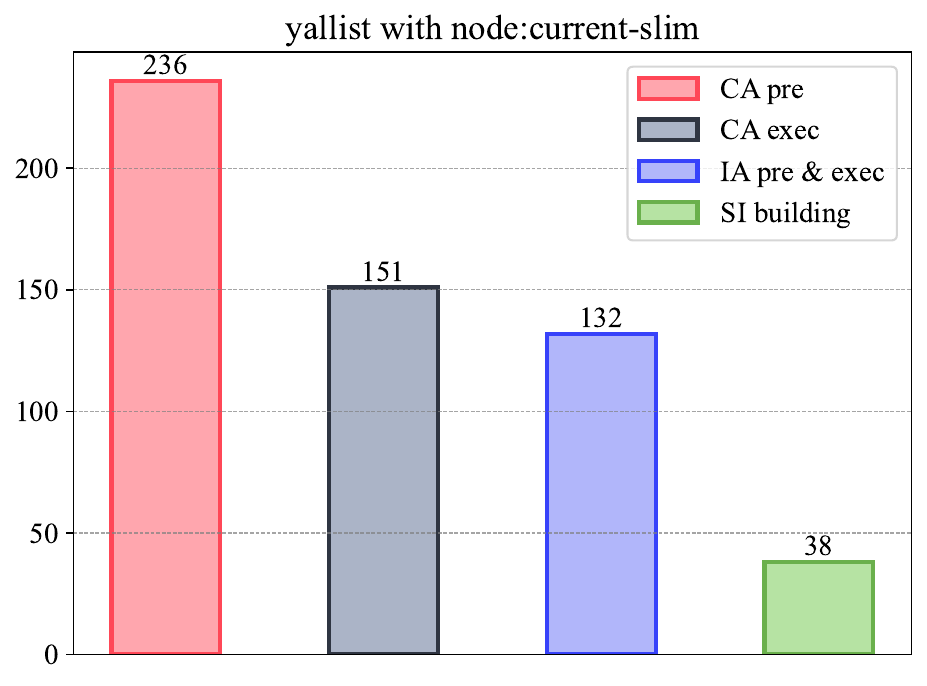}
	}
	\subfigure[estraverse]{
		\centering
		\includegraphics[width=0.18\textwidth]{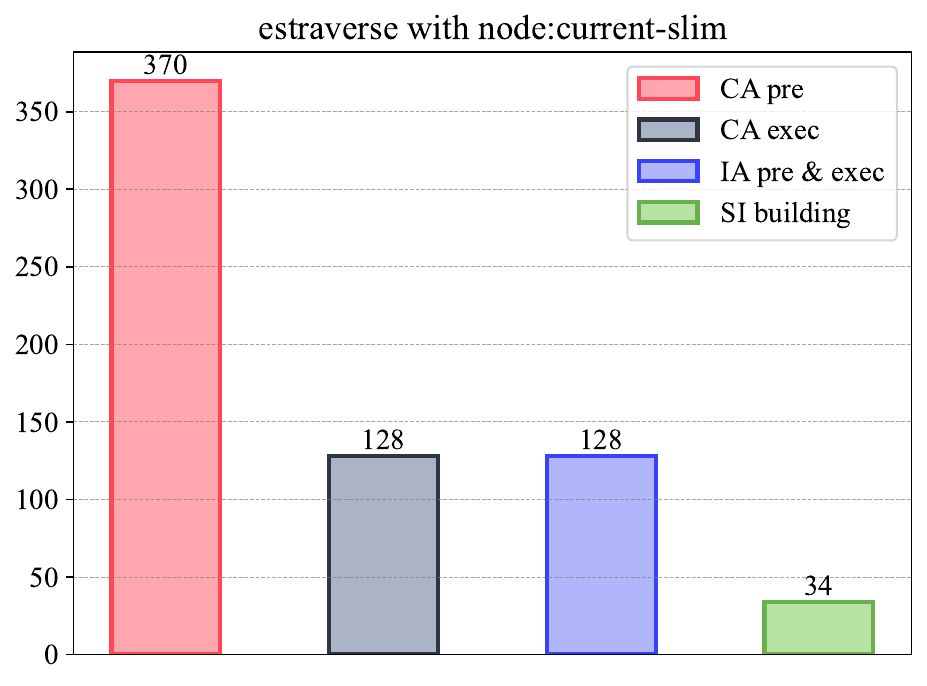}
	}

	\subfigure[deepmerge]{
		\centering
		\includegraphics[width=0.18\textwidth]{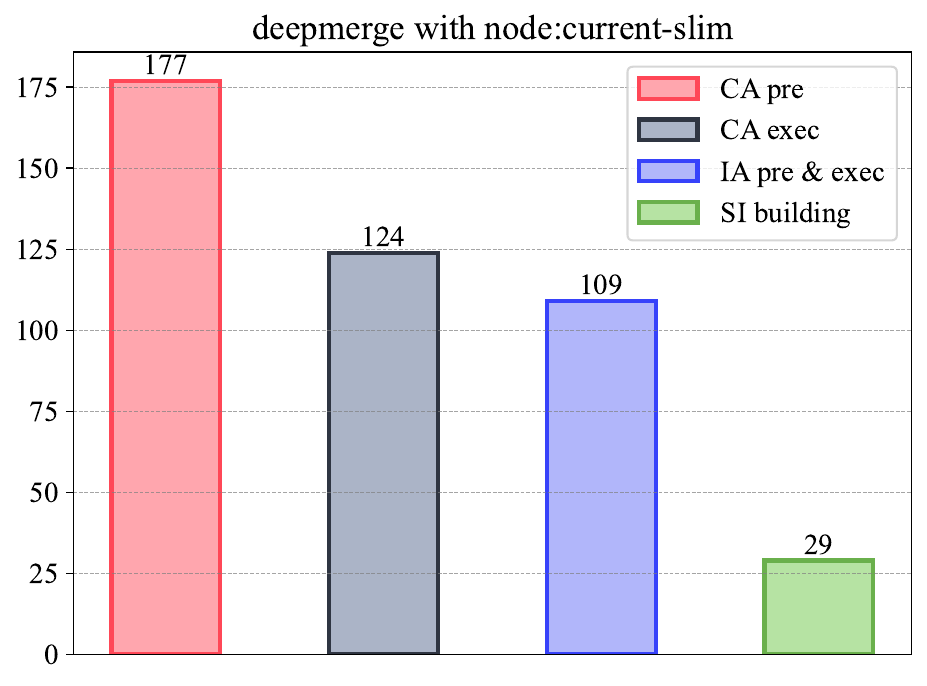}
	}
	\subfigure[node-fs-extra]{
		\centering
		\includegraphics[width=0.18\textwidth]{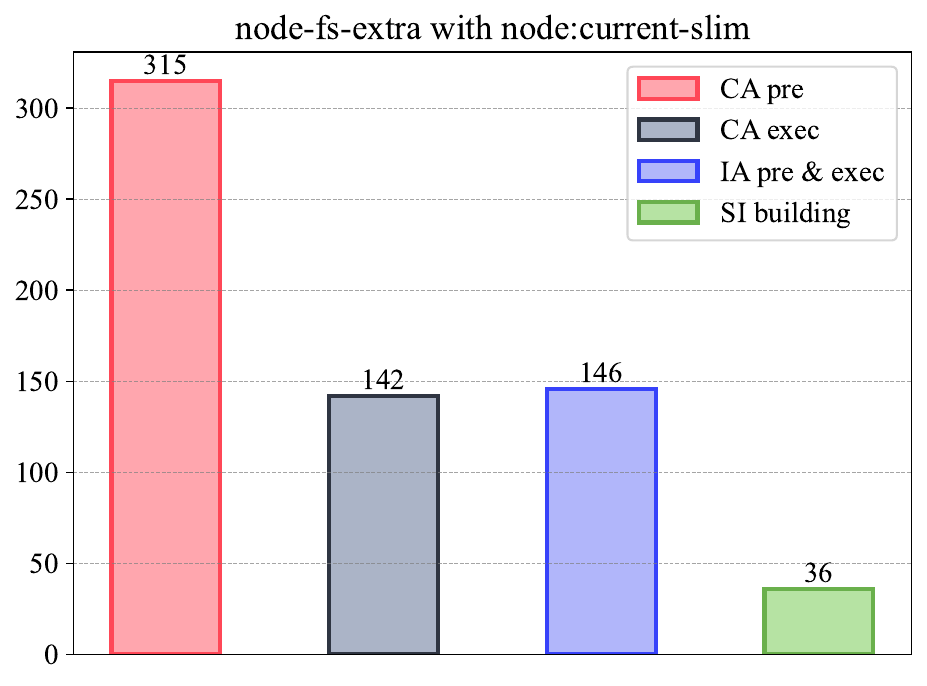}
	}
	\subfigure[node-jsonwebtoken]{
		\centering
		\includegraphics[width=0.185\textwidth]{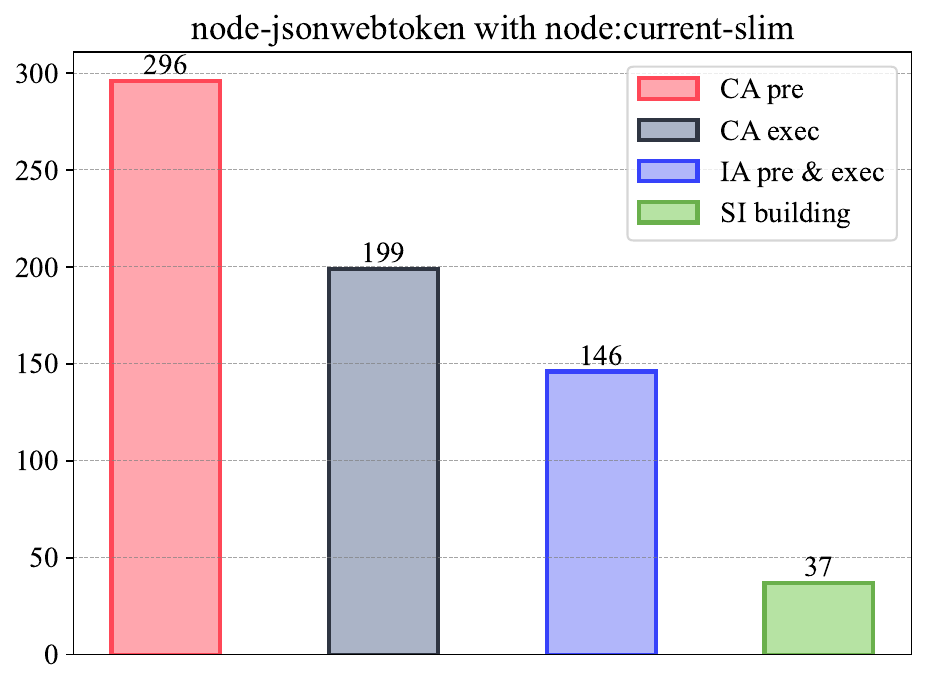}
	}
	\subfigure[node-which]{
		\centering
		\includegraphics[width=0.18\textwidth]{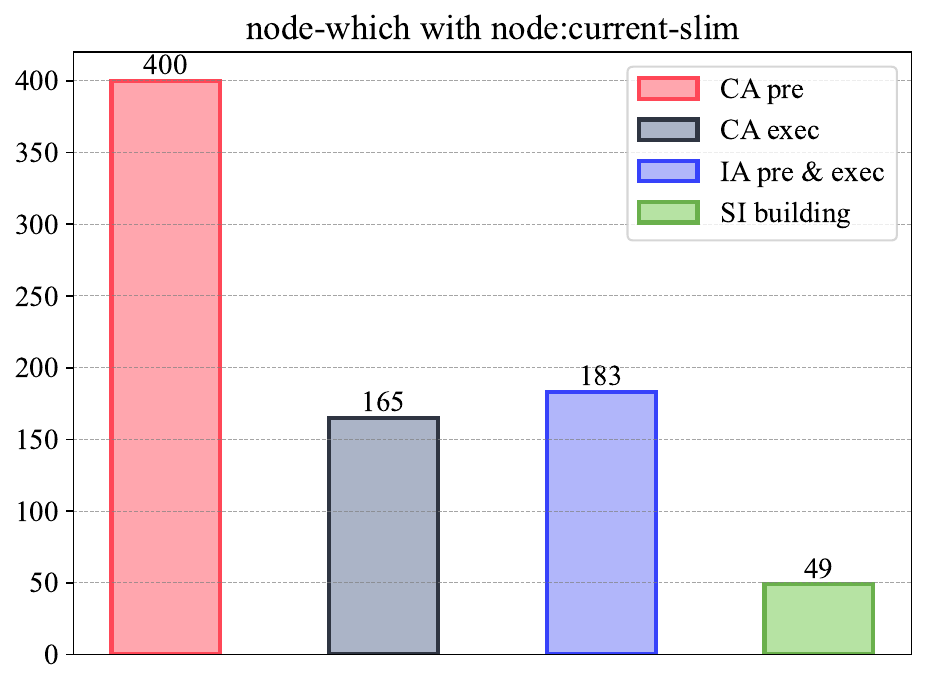}
	}
	\subfigure[prompt]{
		\centering
		\includegraphics[width=0.18\textwidth]{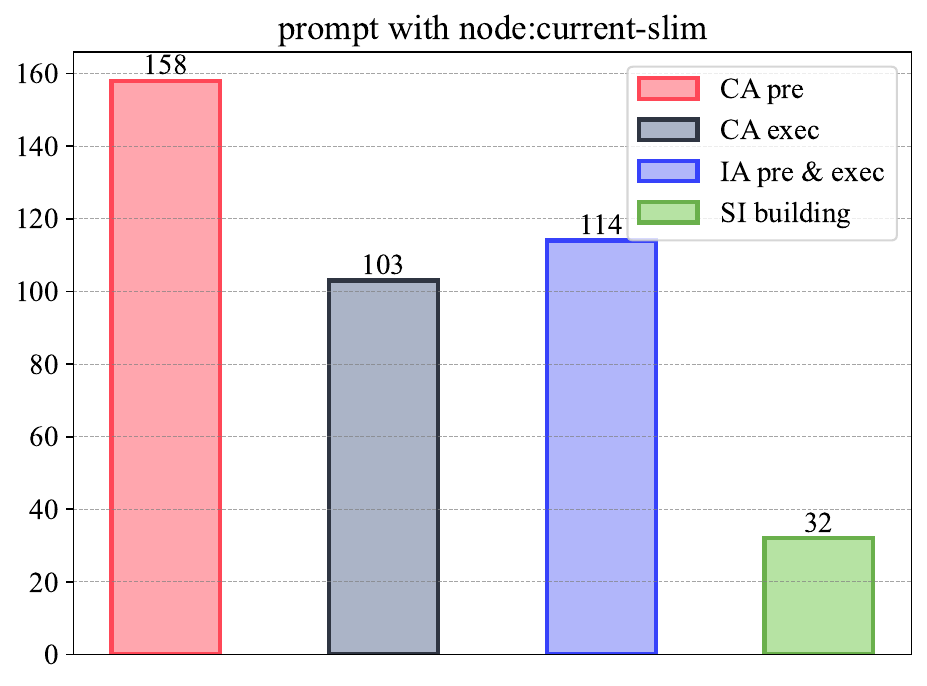}
	}

	\subfigure[shelljs]{
		\centering
		\includegraphics[width=0.18\textwidth]{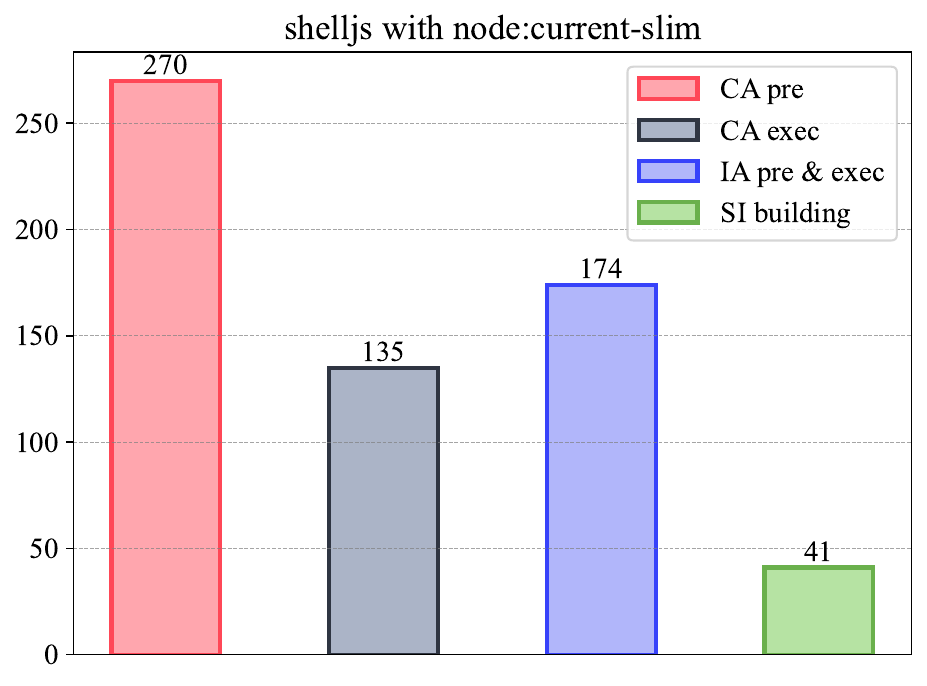}
	}
	\subfigure[winston]{
		\centering
		\includegraphics[width=0.18\textwidth]{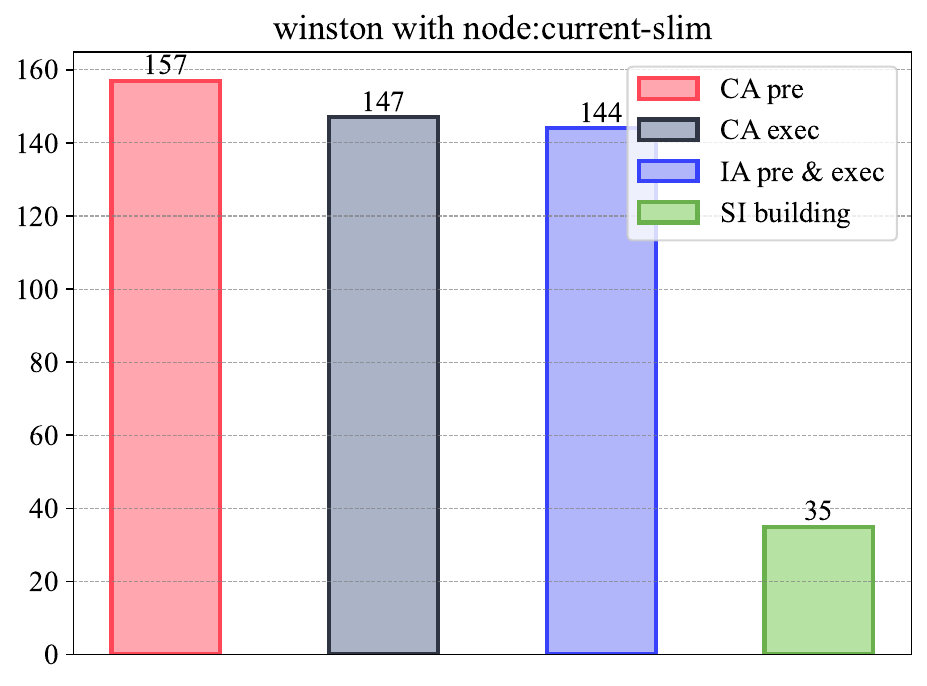}
	}
	\subfigure[ws]{
		\centering
		\includegraphics[width=0.18\textwidth]{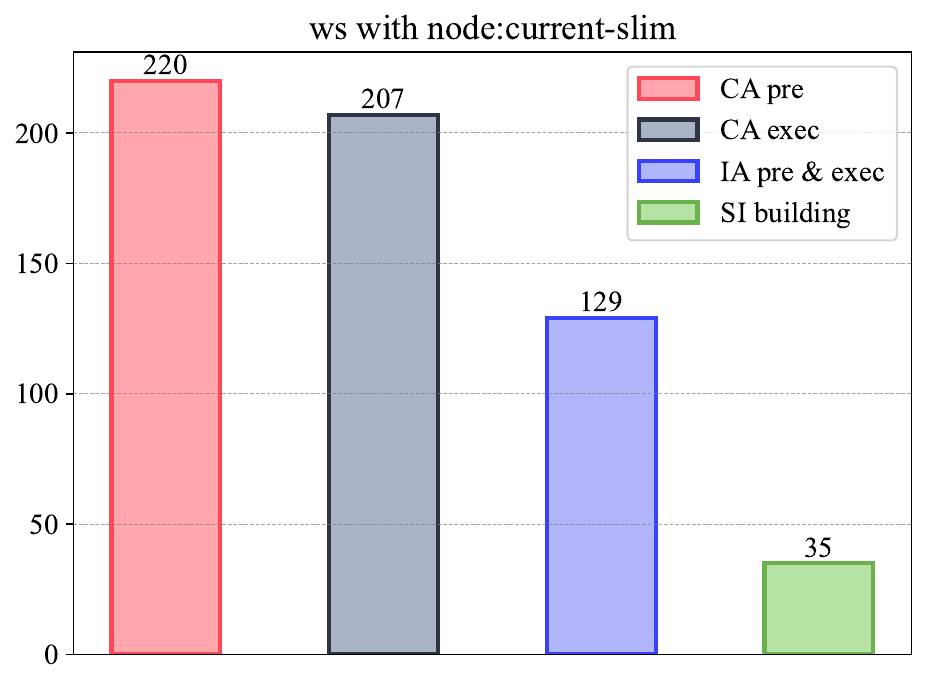}
	}
	\subfigure[minimist]{
		\centering
		\includegraphics[width=0.18\textwidth]{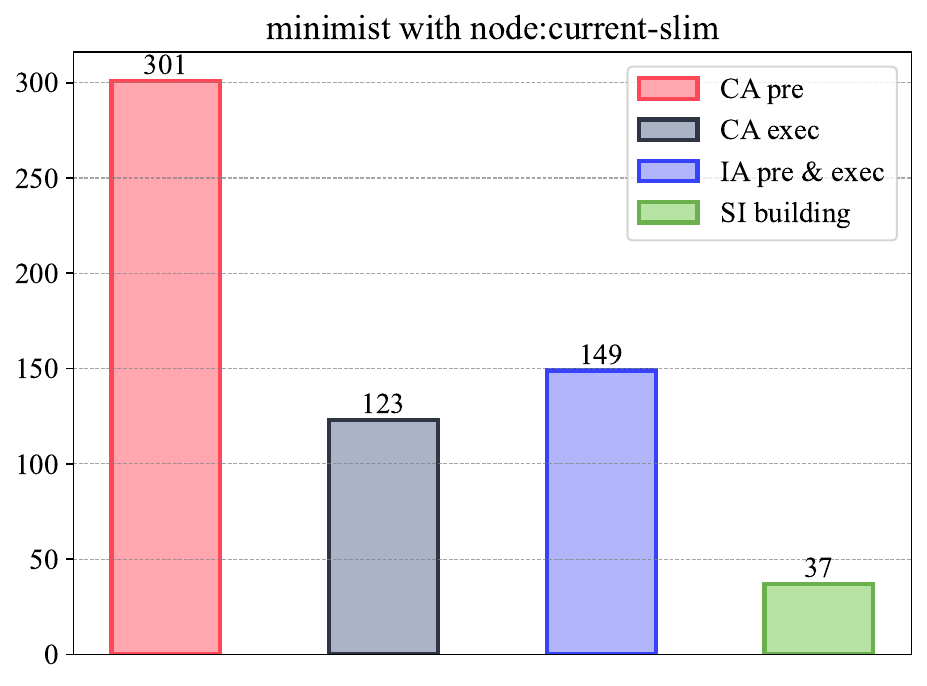}
	}
	\subfigure[node-portfinder]{
		\centering
		\includegraphics[width=0.18\textwidth]{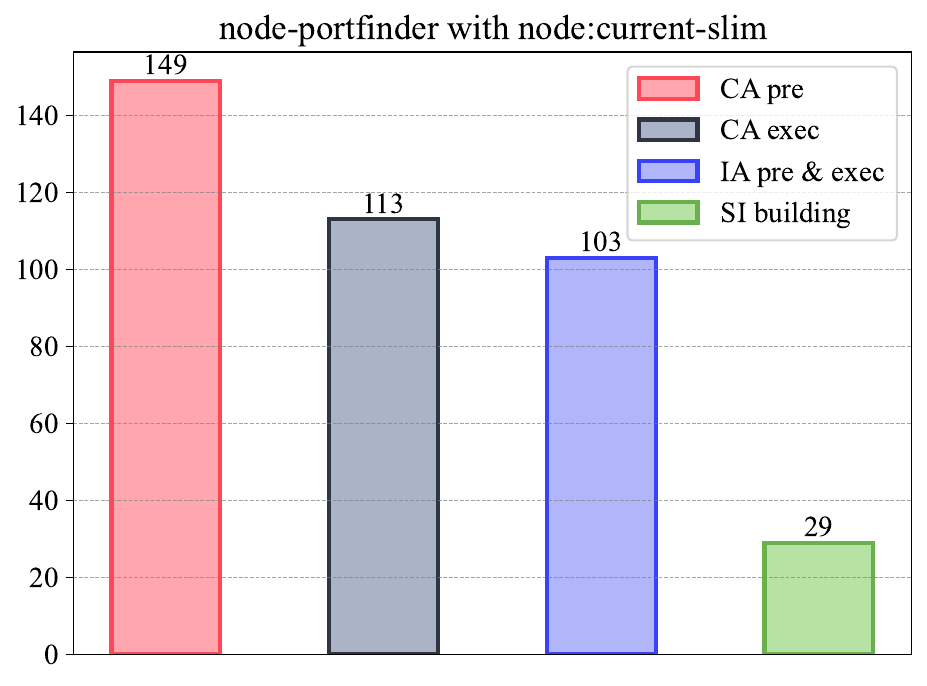}
	}
	
	\caption{Efficiency of $\delta$-SCALPEL when specifying image's entry point. The base image is \texttt{node:current-slim}. CA pre indicates code analyzer preparation, CA exec indicates code analyzer execution, IA pre \& exec indicates image analyzer preparation \& execution, and SI building indicates slim image building. The Y-axis indicates the runtime, expressed in seconds (s).}
	\label{rq2ncs}
\end{figure*}

\begin{figure*}[!t]
	\centering
	\subfigure[semver]{
		\centering
		\includegraphics[width=0.18\textwidth]{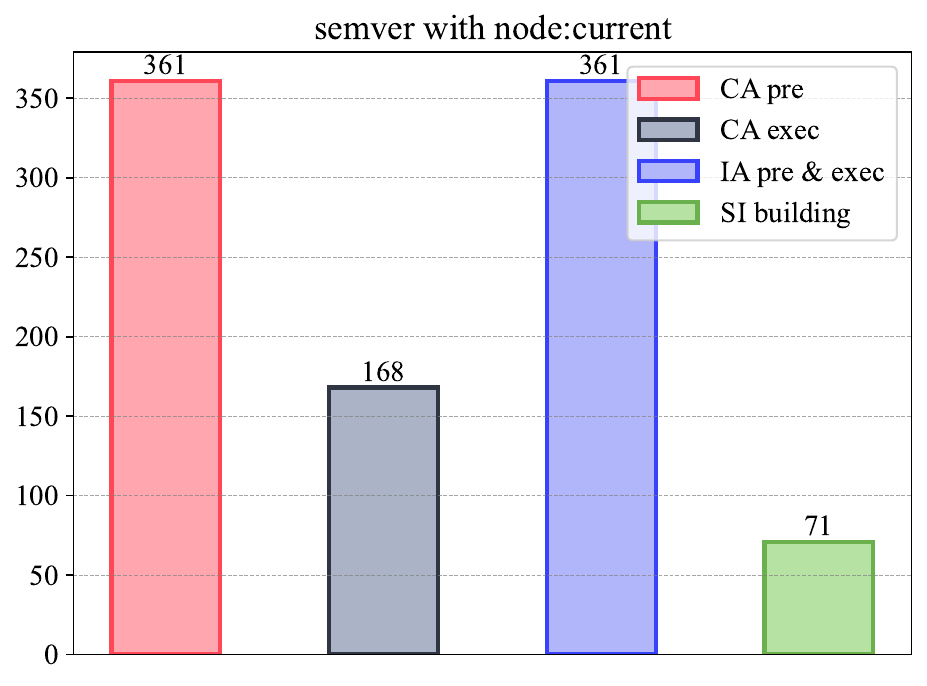}
	}
	\subfigure[chalk]{
		\centering
		\includegraphics[width=0.18\textwidth]{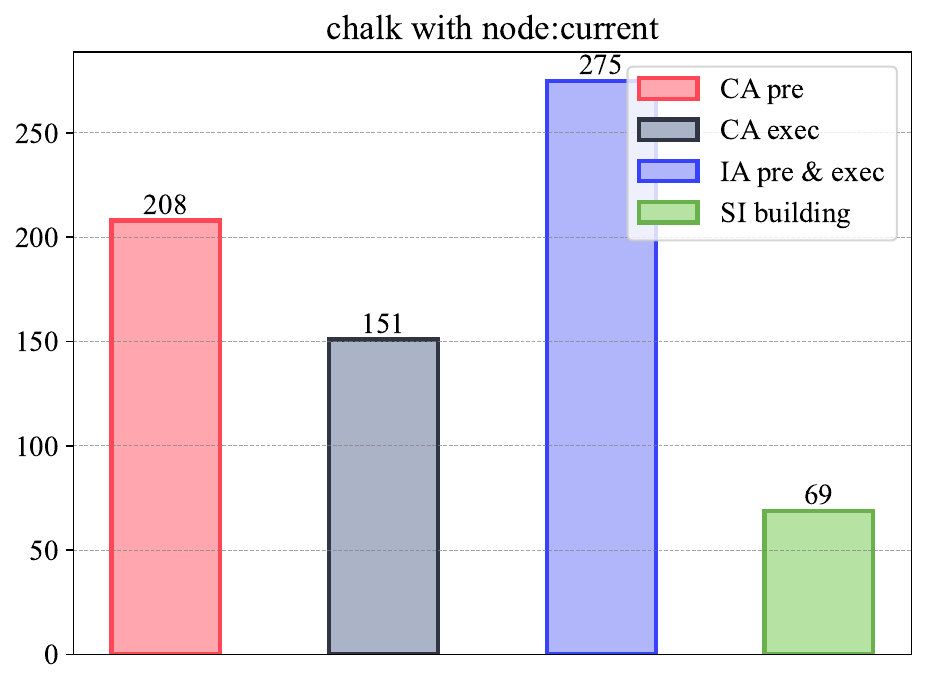}
	}
	\subfigure[nodejs-websocket]{
		\centering
		\includegraphics[width=0.18\textwidth]{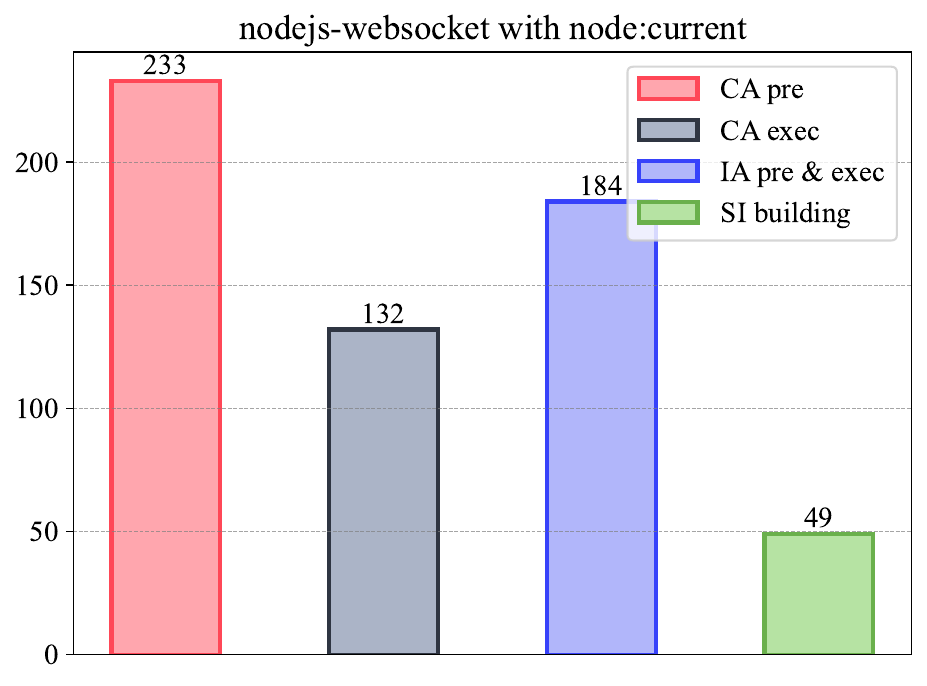}
	}
	\subfigure[lru-cache]{
		\centering
		\includegraphics[width=0.18\textwidth]{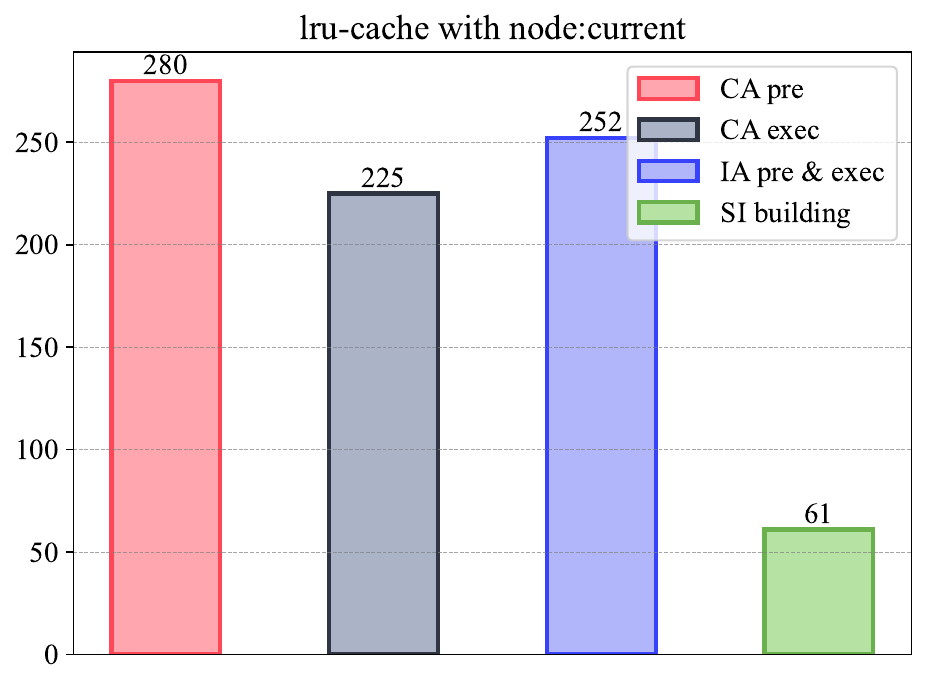}
	}
	\subfigure[minimatch]{
		\centering
		\includegraphics[width=0.18\textwidth]{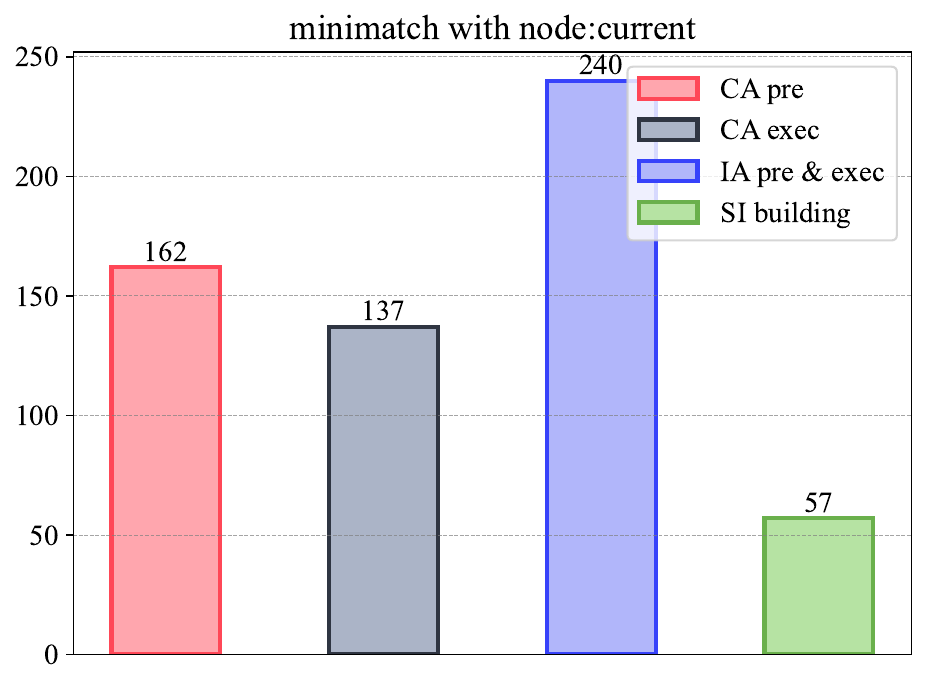}
	}
	
	\subfigure[strip-ansi]{
		\centering
		\includegraphics[width=0.18\textwidth]{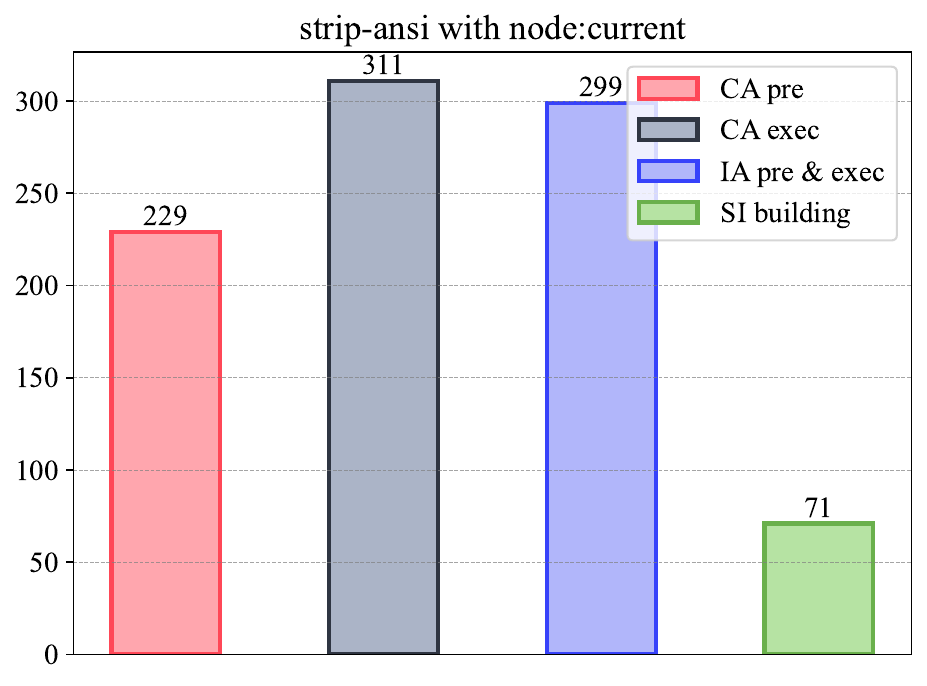}
	}
	\subfigure[node-glob]{
		\centering
		\includegraphics[width=0.18\textwidth]{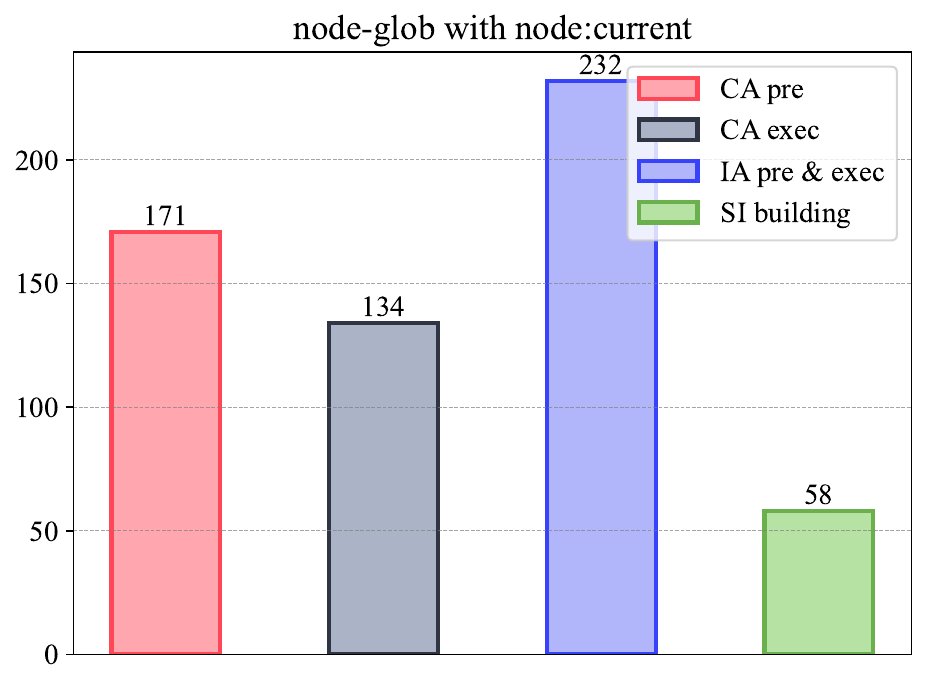}
	}
	\subfigure[commander.js]{
		\centering
		\includegraphics[width=0.18\textwidth]{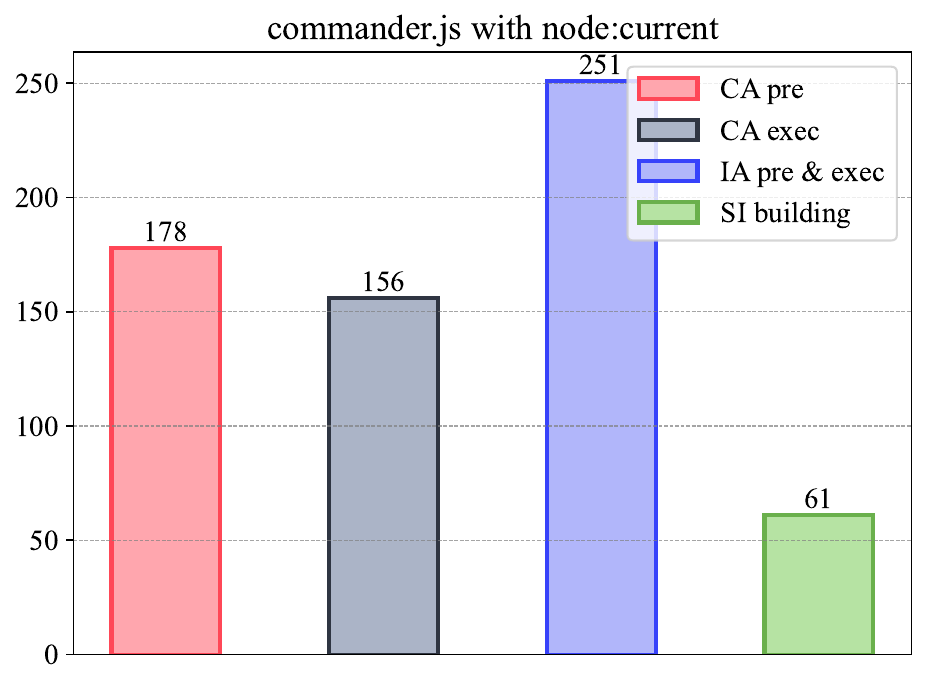}
	}
	\subfigure[yallist]{
		\centering
		\includegraphics[width=0.18\textwidth]{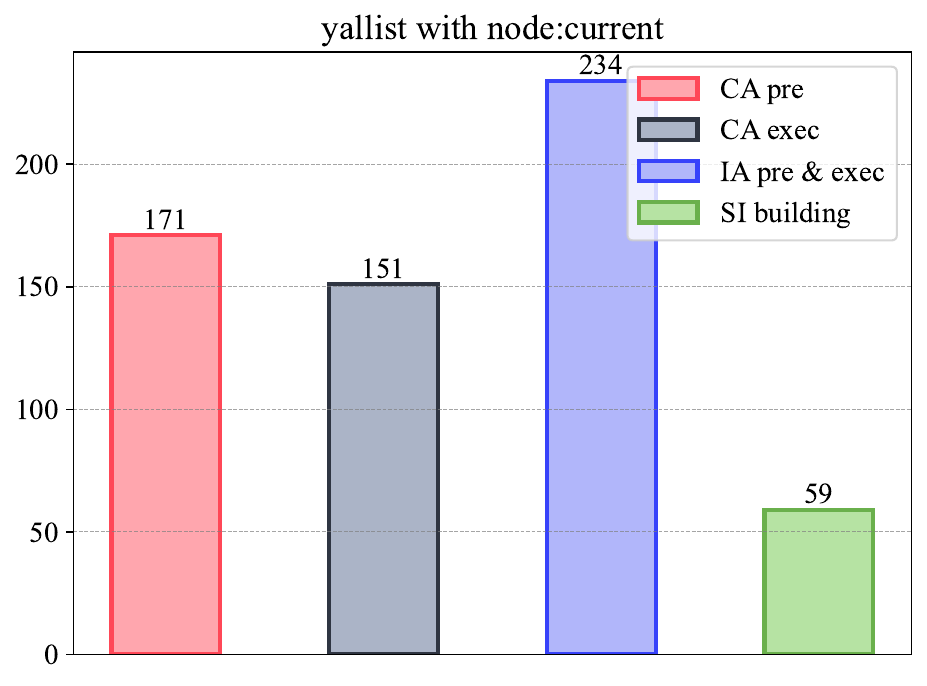}
	}
	\subfigure[estraverse]{
		\centering
		\includegraphics[width=0.18\textwidth]{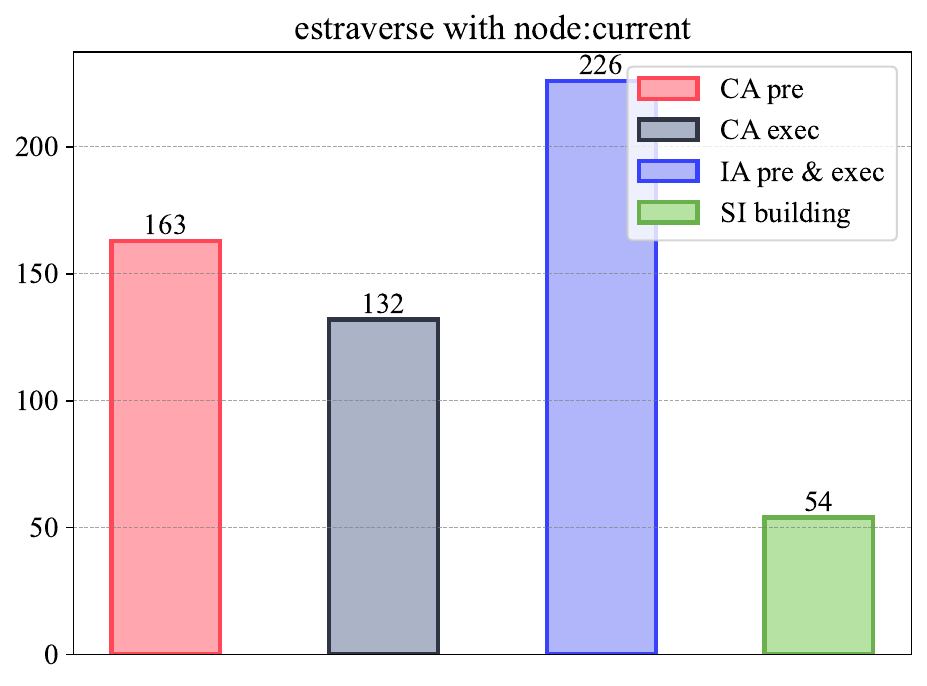}
	}
	
	\subfigure[deepmerge]{
		\centering
		\includegraphics[width=0.18\textwidth]{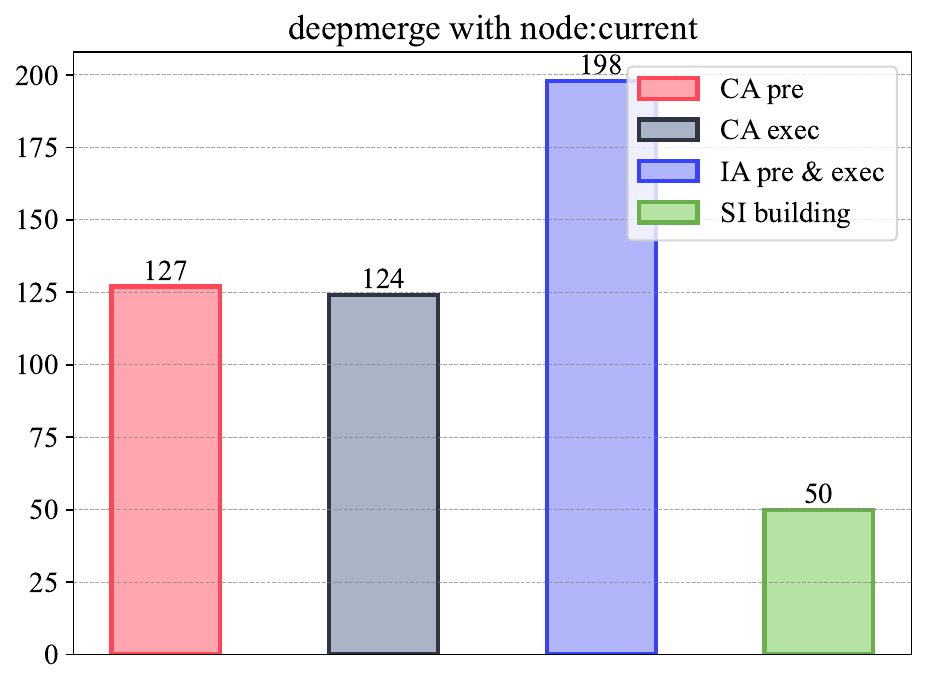}
	}
	\subfigure[node-fs-extra]{
		\centering
		\includegraphics[width=0.18\textwidth]{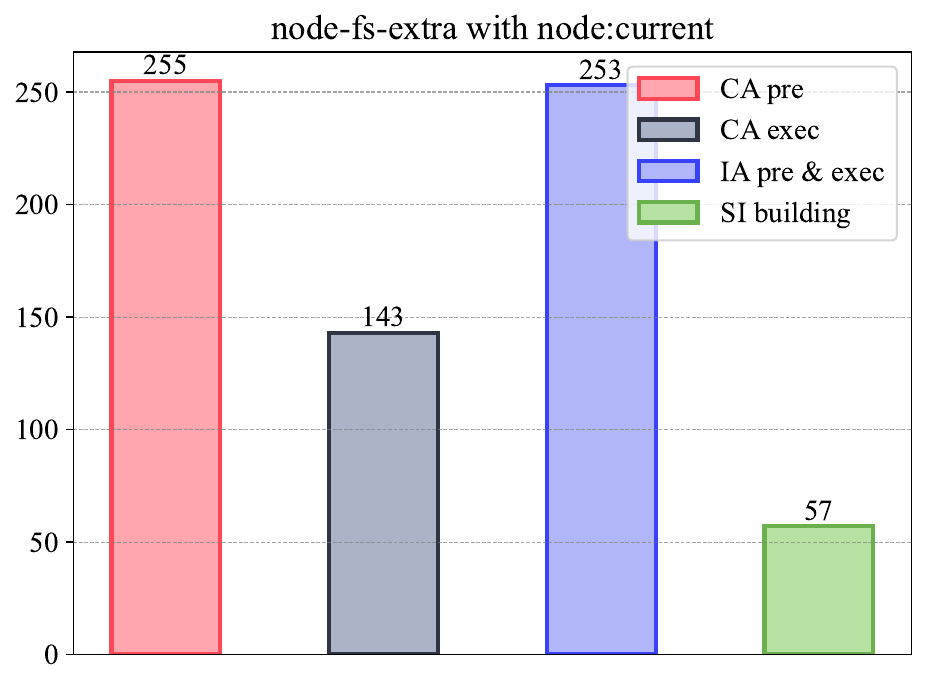}
	}
	\subfigure[node-jsonwebtoken]{
		\centering
		\includegraphics[width=0.185\textwidth]{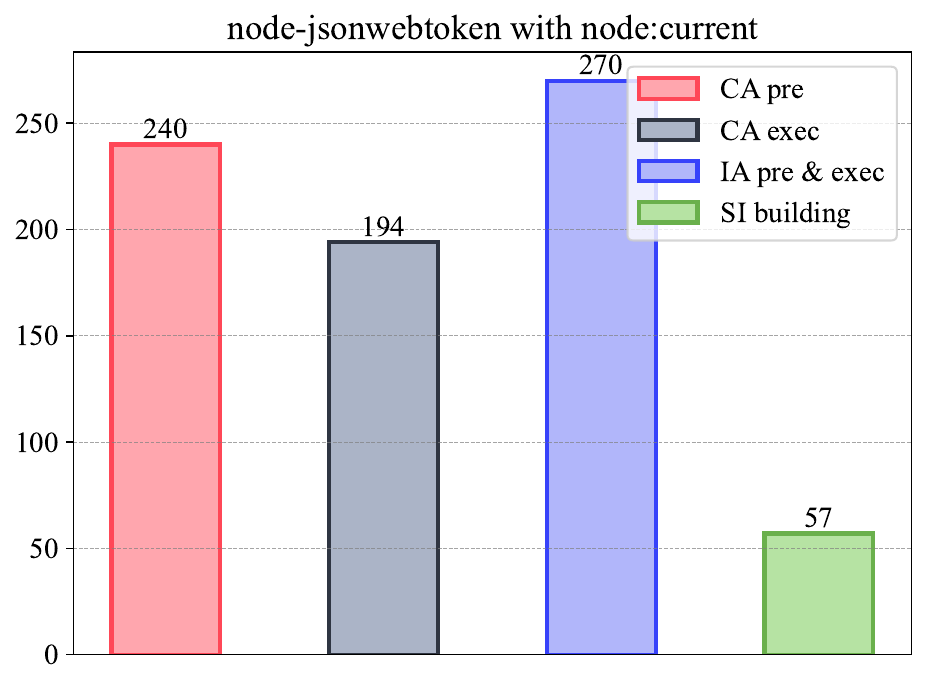}
	}
	\subfigure[node-which]{
		\centering
		\includegraphics[width=0.18\textwidth]{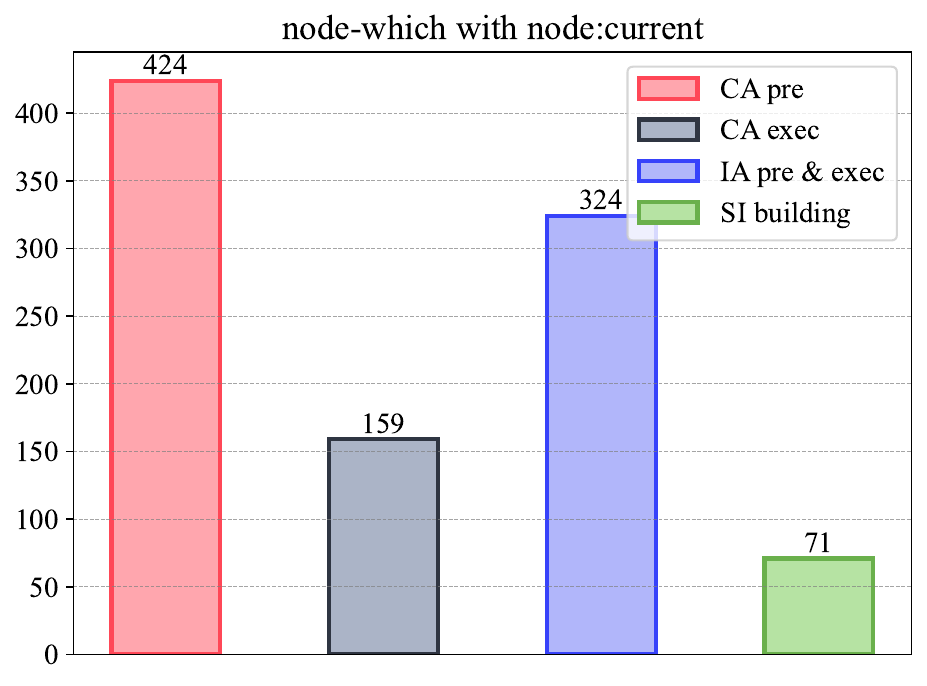}
	}
	\subfigure[prompt]{
		\centering
		\includegraphics[width=0.18\textwidth]{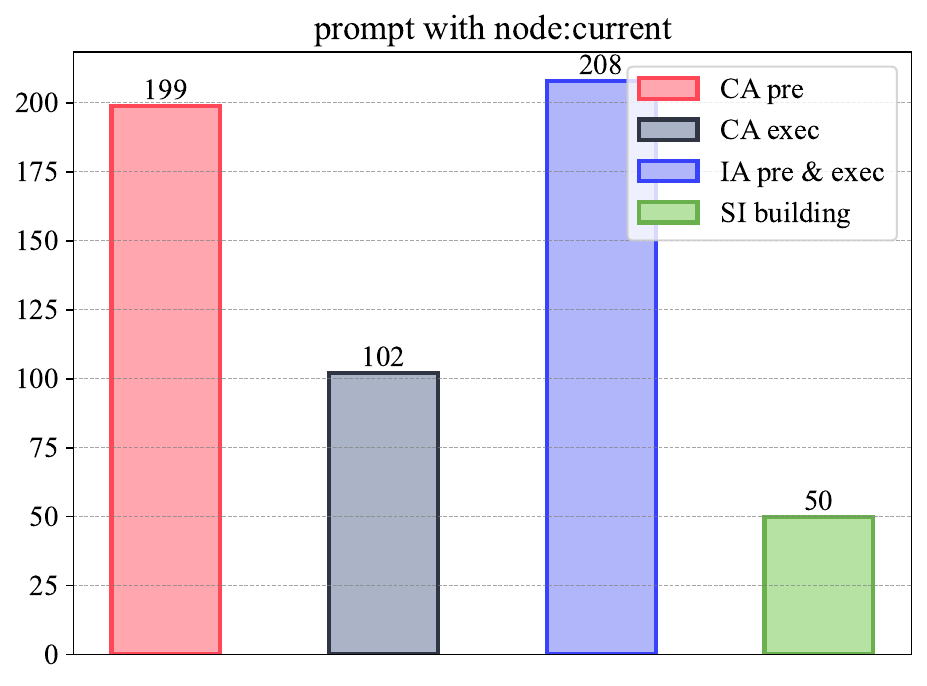}
	}
	
	\subfigure[shelljs]{
		\centering
		\includegraphics[width=0.18\textwidth]{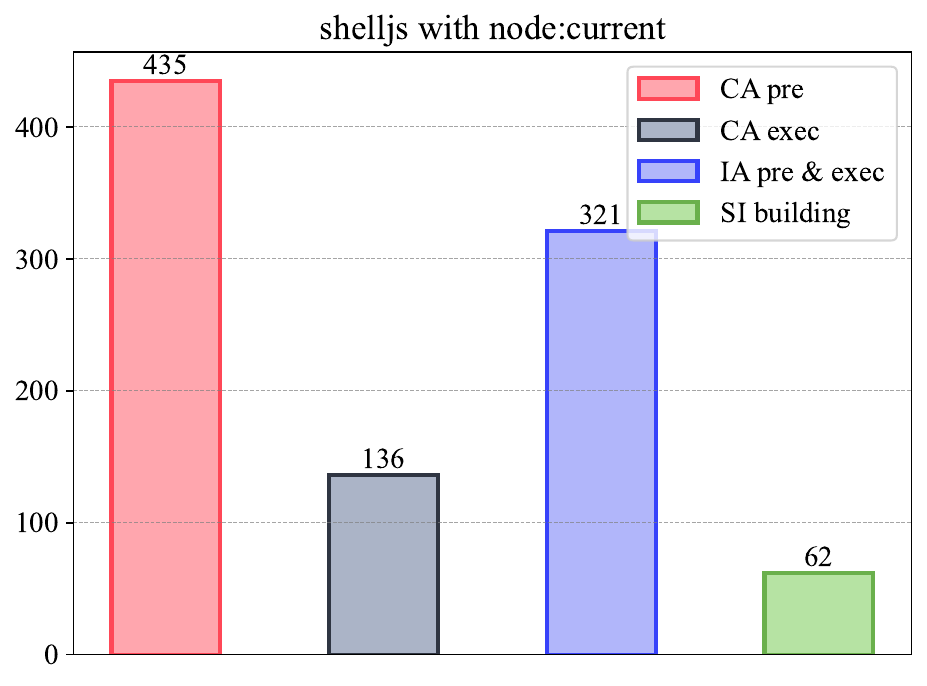}
	}
	\subfigure[winston]{
		\centering
		\includegraphics[width=0.18\textwidth]{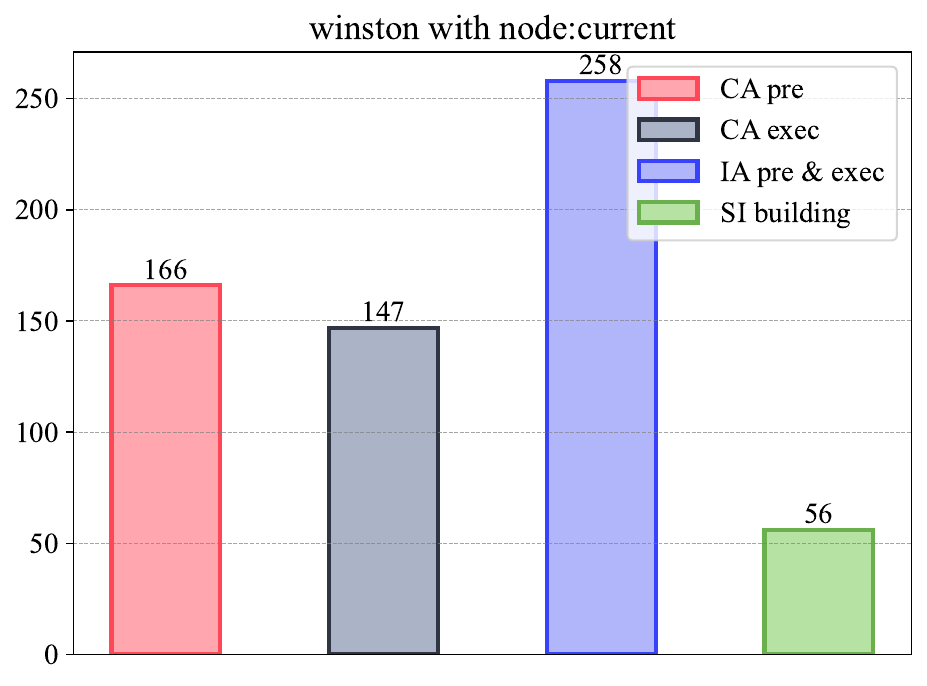}
	}
	\subfigure[ws]{
		\centering
		\includegraphics[width=0.18\textwidth]{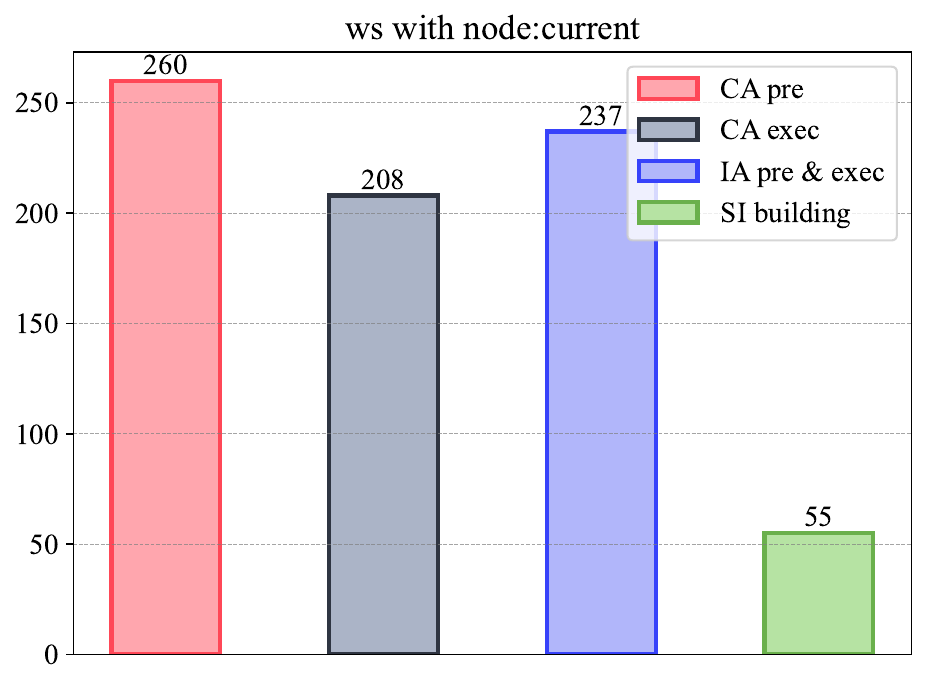}
	}
	\subfigure[minimist]{
		\centering
		\includegraphics[width=0.18\textwidth]{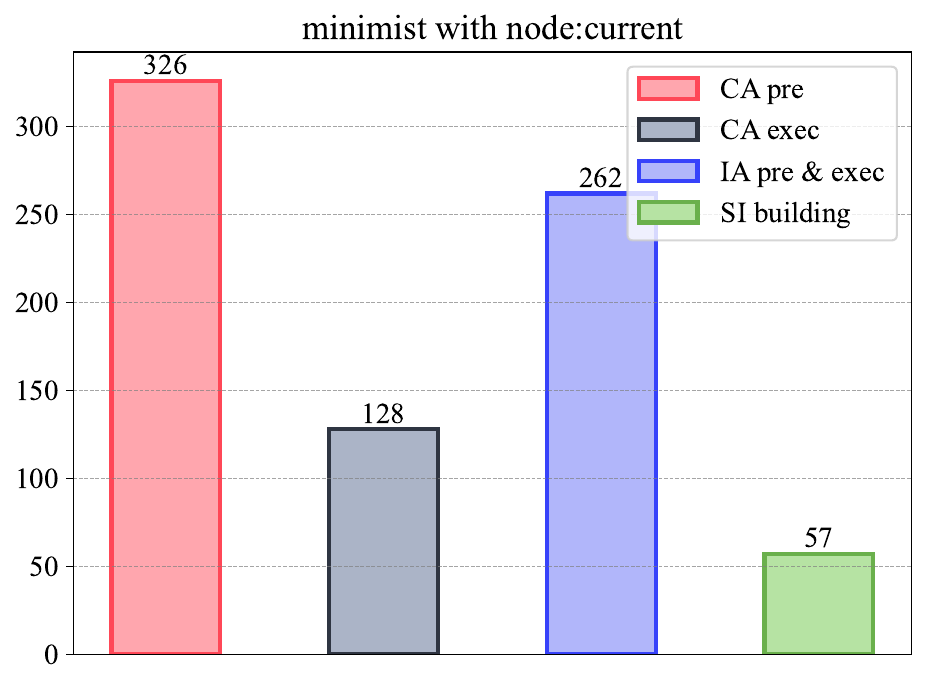}
	}
	\subfigure[node-portfinder]{
		\centering
		\includegraphics[width=0.18\textwidth]{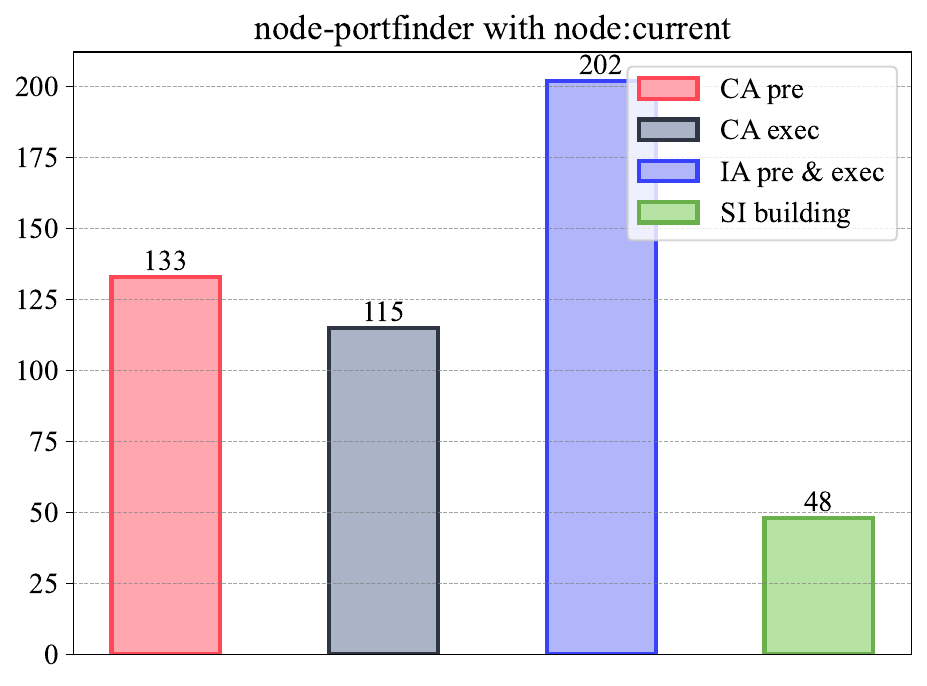}
	}
	
	\caption{Efficiency of $\delta$-SCALPEL when specifying image's entry point. The base image is \texttt{node:current}. CA pre indicates code analyzer preparation, CA exec indicates code analyzer execution, IA pre \& exec indicates image analyzer preparation \& execution, and SI building indicates slim image building. The Y-axis indicates the runtime, expressed in seconds (s).}
	\label{rq2nc}
\end{figure*}

Evaluation results are shown in Fig. \ref{rq2ncs} and Fig. \ref{rq2nc}. Generally, the code analyzer preparation part is time-consuming. In this part, $\delta$-SCALPEL configures the image for code analyzer, which includes setting up the base image, configuring the CodeQL environment, and downloading and installing NPM dependencies for the project. Due to network limitations, the majority of the time is spent downloading the base image and NPM dependencies, resulting in an extended preparation time for the code analyzer. In the code analyzer execution part, it takes over 100 seconds to complete the process (with the longest analysis time being 311 seconds for the strip-ansi image based on \texttt{node:current}). This extended time is due to the need to analyze the data dependencies of both the project code and the packages it depends on. The long dependency chain of the project results in a significant amount of source code that must be analyzed. 

Different base images have minimal impact on the execution efficiency of the code analyzer, as it only analyzes the project code and the packages it depends on, regardless of the base image. In contrast, the size of the base image affects the runtime of the image analyzer. When the base image is switched from \texttt{node:current-slim} to \texttt{node:current}, the root file system content increases, resulting in more data for the image analyzer to process, which extends the runtime. This also increases the size of the \texttt{rootfs.tar} file generated by the image analyzer, which in turn affects the building time of the slim image.

\begin{mybox}
	\textbf{Answer to RQ2:} Due to network limitations, the code analyzer preparation part is time-consuming. This analyzer must process both the project code and its NPM dependencies, resulting in a runtime of over 100 seconds, though this is unaffected by the base image size. However, the runtime of the image analyzer and the slim image building parts increases as the size of the base image grows.
\end{mybox}

\subsection{RQ3: Significance}

The primary goal of image slimming is to reduce the image size, improving the ease of transmission and storage. An indirect goal is to minimize the attack surface of the image. Identifying and removing unnecessary code fragments and components is a key approach to reducing the attack surface \cite{mishra2018shredder,mulliner2015breaking,quach2018debloating,koo2019configuration}. In the Application Container Security Guide published by NIST \cite{nistSP800190}, it is recommended to limit the available functions of containers to reduce the attack surface. In this research question, we assess the size of the image's attack surface by measuring the number of commands supported before and after image slimming. Specifically, we use the selected 20 NPM projects to build images based on \texttt{node:current-slim} and \texttt{node:current}, respectively. We then use $\delta$-SCALPEL to slim these images and calculate the average number of commands supported by them before and after slimming.

\begin{figure}[!htpb]
	\centering
	\subfigure[node:current-slim]{
		\centering
		\includegraphics[width=0.17\textwidth]{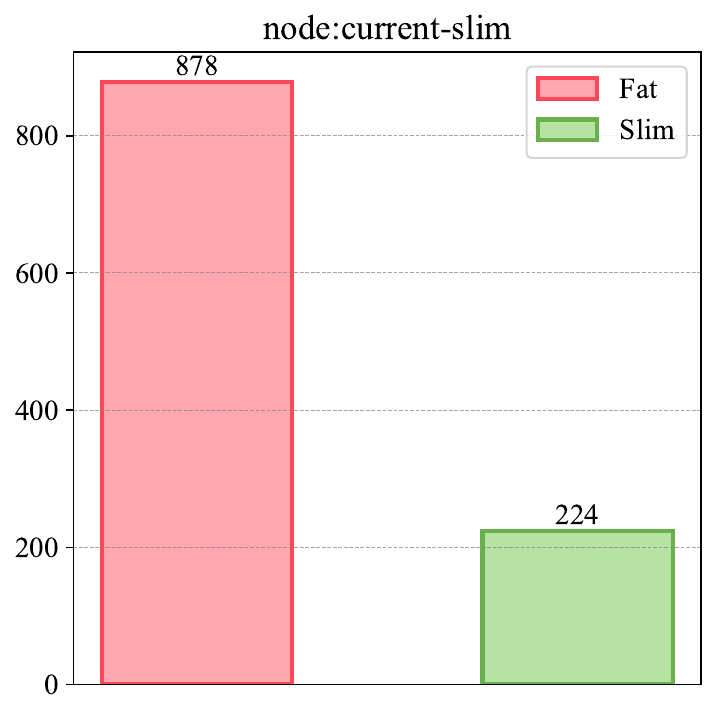}
	}
	\hspace{30pt}
	\subfigure[node:current]{
		\centering
		\includegraphics[width=0.17\textwidth]{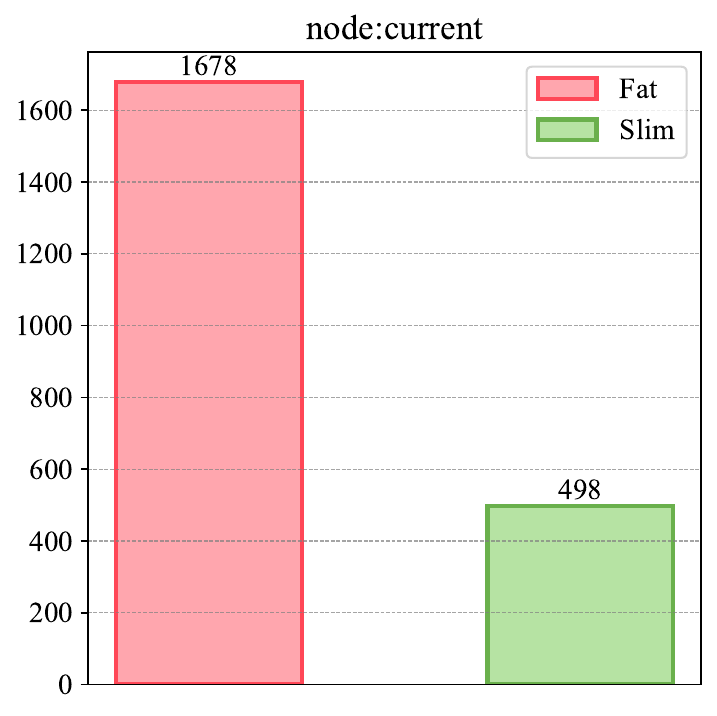}
	}
	
	\caption{Comparison of the average number of commands supported by the images based on \texttt{node:current-slim} and \texttt{node:current}, both before and after the slimming process. The Y-axis indicates the number of commands supported by the image.}
	\label{rq3}
\end{figure}

The evaluation results are shown in Fig. \ref{rq3}. \texttt{node:current-slim} is a condensed version of \texttt{node:current}, containing only the minimal packages required to run Node.js \cite{dockerNodeGitHub}. As shown in Fig. \ref{rq3}(a), \texttt{node:current} supports 1,678 commands, whereas in Fig. \ref{rq3}(b), \texttt{node:current-slim} supports 878 commands, which is 800 fewer than \texttt{node:current}. That is to say, without using $\delta$-SCALPEL, the attack surface of the image based on \texttt{node:current-slim} is reduced by 47.7\% compared to the image based on \texttt{node:current}. For images based on \texttt{node:current-slim}, the average number of supported commands decreases to 224 after slimming with $\delta$-SCALPEL, leading to a 74.5\% reduction in the attack surface compared to the pre-slimming state. In comparison, for images based on \texttt{node:current}, the average number of supported commands is 498, resulting in a 70.3\% decrease in the attack surface relative to its original state.

\begin{mybox}
	\textbf{Answer to RQ3:} The attack surface of the official slim image is 47.7\% smaller than that of the standard image. After applying $\delta$-SCALPEL, the attack surface is reduced by 74.5\% for images based on \texttt{node:current-slim}, while for images based on \texttt{node:current}, the attack surface is reduced by 70.3\%.
\end{mybox}

\subsection{Threats to Validity}

In this paper, we propose the use of static data dependency analysis to extract the environment dependencies of the project code, aiming to reduce the image size while ensuring the normal operation of the project code. The evaluation results demonstrate that the proposed $\delta$-SCALPEL model is effective. However, $\delta$-SCALPEL still faces the following limitations:

\noindent\textbf{\uline{Threats to effectiveness.}} From the evaluation results shown in Tab. \ref{rq1.1} and Tab. \ref{rq1.2}, it is evident that $\delta$-SCALPEL can achieve precise and effective image slimming, regardless of whether the image entry point is explicitly specified. In this paper, we focus on removing executable files in the \texttt{/bin} and \texttt{/sbin} folders and dynamic link library files in the \texttt{/lib} folder, as these files are closely related to the execution of the project code. The base runtime environment of the container, including files in the \texttt{/var} and \texttt{/opt} directories, is fully retained. While this ensures the smooth operation of the container and project code, it inevitably reduces the slimming rate. Therefore, in future work, we will further analyze these files and identify the minimal environment dependencies required for container operation by incorporating dynamic analysis techniques. This will enhance $\delta$-SCALPEL's image slimming capabilities.

\noindent\textbf{\uline{Threats to efficiency.}} Besides effectiveness, efficiency is a crucial evaluation metric to determine whether a model can be applied in the real production environment. As shown in the evaluation results in Fig. \ref{rq2ncs} and Fig. \ref{rq2nc}, the average execution time of $\delta$-SCALPEL is approximately ten minutes, with the most time-consuming parts being the preparation of the image for the code analyzer and the execution of the code analyzer itself. The running time for the image preparation part of the code analyzer is constrained by the local network environment, as it takes considerable time to download the base image and NPM packages. Despite optimizing the static data dependency analysis algorithm for the code analyzer, we still encounter challenges of efficiency. Therefore, in our future research, we will enhance the efficiency of the code analyzer by incorporating more advanced analysis algorithms and designing more effective package filtering mechanisms.

\section{Conclusion}

Image slimming is a valuable area of research that can help reduce the image size and minimize the attack surface of containers. Existing method determines a project's environment dependencies by observing the operational behavior of the container. However, this approach has significant limitations, including the incomplete extraction of the environment dependencies and failure to slim the image, especially when the image's entry point is not specified. In this paper, we propose $\delta$-SCALPEL, a model that utilizes static code analysis technology to extract the environment dependencies of a project, thereby addressing the limitations of the existing method. Our evaluation results demonstrate that $\delta$-SCALPEL effectively reduces the image size while ensuring the normal operation of the project.

\section*{Data Availability Statement}

The replication package of $\delta$-SCALPEL is available at \url{https://zenodo.org/records/13982576}.

	


\bibliographystyle{ACM-Reference-Format}
\bibliography{sample-acmsmall-conf}

\end{document}